\documentclass[a4paper, draftcls, onecolumn, romanappendices, 12pt]{IEEEtran}

\usepackage[dvipsnames]{xcolor}
\usepackage[nolist]{acronym}
\usepackage{bm}
\usepackage{mathrsfs}
\usepackage{amsmath,amsbsy,amsfonts,amssymb}
\usepackage{subcaption}
\usepackage{mathtools}

\mathtoolsset{showonlyrefs}

\newcounter{example}
\newtheorem{Example}[example]{Example}

\newcommand{\IAPP}{\mathsf{I}_{\mathsf{APP}}}
\newcommand{\load}{\mathsf{G}}
\newcommand{\plr}{\mathsf{p}_l}

\newcommand{\snr}{\ensuremath{E_s/N_0}}
\newcommand{\sLen}{t_s}
\newcommand{\ns}{n_s}
\newcommand{\rx}{y}
\newcommand{\rxVec}{\bm{\rx}}
\newcommand{\tx}{x}

\newcommand{\noisev}{n}
\newcommand{\noisePr}{\nu}
\newcommand{\noiseVec}{\bm{\noisev}}
\newcommand{\noiseSD}{N_0}
\newcommand{\intVal}{z}
\newcommand{\intVec}{\bm{\intVal}}

\newcommand{\symVal}{a}
\newcommand{\pulse}{g}

\newcommand{\rep}{r}
\newcommand{\repInd}{\bar \rep}
\newcommand{\rSet}{\mathscr{R}}
\newcommand{\sPw}{\mathsf{P}}
\newcommand{\iPw}{\mathsf{Z}}

\newcommand{\tm}{t}
\newcommand{\epoch}{\epsilon}
\newcommand{\Depoch}{\Delta \epoch}
\newcommand{\freq}{f}

\newcommand{\phase}{\varphi}
\newcommand{\Dphase}{\Delta \phase}
\newcommand{\basem}[1]{\mathbf B_{\mathsf{#1}}}

\newcommand{\wind}{W}
\newcommand{\Dwind}{\Delta \wind}

\newcommand{\pkLen}{t_p}
\newcommand{\vf}{t_f}
\newcommand{\dg}{d}

\newcommand{\noise}{\sigma_n^2}
\newcommand{\inter}{\sigma_\iota^2}

\newcommand{\interCap}{\sigma_{\iota,\mathsf{o}}^2}
\newcommand{\nPI}{\sigma^2}
\newcommand{\BnPI}{\bm\sigma^{(2)}}
\newcommand{\BnPIo}{\bm\sigma^{(2)}_{\mathsf p}}

\newcommand{\rvDec}{\mathcal{D}}
\newcommand{\Ca}{\mathsf{C}}
\newcommand{\Capa}{\Ca_\mathsf{o}}
\newcommand{\rate}{\mathsf{R}}

\newcommand{\idt}{\sigma_{\iota,\mathsf{th}}^2}

\newcommand{\alphal}{\alpha^{(\ell)}}
\newcommand{\capBO}{\beta}

\title{Protograph LDPC Code Design for Asynchronous Random Access}
\author{
\vspace{3mm}
\normalsize{Federico Clazzer, Bal\'azs Matuz, Sachini Jayasooriya, Mahyar Shirvanimoghaddam\\ and Sarah J. Johnson}
\thanks{F. Clazzer and B. Matuz are with the Inst.\ of Communications and Navigation, German Aerospace Center (DLR), Oberpfaffenhofen, Wessling, Germany (e-mail: \{federico.clazzer,balazs.matuz\}@dlr.de),\newline
S. Jayasooriya and S. J. Johnson are with the School of Electrical Engineering and Computing, The University of Newcastle, University Dr, Callaghan NSW 2308 Newcastle, Australia (e-mail: \{sarah.johnson,sachini.jayasooriya\}@newcastle.edu.au),\newline
M. Shirvanimoghaddam is with the Faculty of Engineering, The University of Sydney, NSW 2006 Sydney, Australia (e-mail: mahyar.shirvanimoghaddam@sydney.edu.au) \newline
Part of this work has been presented at the IEEE 10th International Symposium on Turbo Codes Iterative Information Processing (ISTC), 2018, Hong Kong.
}
}

\begin{document}

\setcounter{page}{1}
\maketitle
\thispagestyle{empty}

% Abstract (Do not insert blank lines, i.e. \\)
\begin{abstract}
This work addresses the physical layer channel code design for an uncoordinated, frame- and slot-asynchronous random access protocol. Starting from the observation that collisions between two users yield very specific  interference patterns, we define a surrogate channel model and propose different protograph low-density parity-check code designs. The proposed codes are both tested in a setup where the physical layer is abstracted, as well as on a more realistic channel model, where finite-length physical layer simulations of the entire asynchronous random access scheme, including decoding are carried out. We find that the abstracted physical layer model overestimates the performance when short blocks are considered. Additionally, the optimized codes show gains in supported channel traffic \--- a measure of the number of terminals that can be concurrently accommodated on the channel \--- of around 17\% at a packet loss rate of $10^{-2}$ w.r.t.\ off-the-shelf codes.
\end{abstract}

%\IEEEpeerreviewmaketitle
%%%%%%%%%%%%%%%%%%%%%%%%%%%%%%%%%%%%%%%%%%
%%%%%%%%%%%%%%%%%%%%%%%%%%%%%%%%%%%%%%%%%%%%%%%%%%%%%%%%%%%%%%%%%%%
%%%
%%% abbreviations.bib
%%%
%%% \acresetall resets all acronyms to not used. Useful after the abstract to redefine all acronyms in the introduction.
%%% \acf{label} Full: written out form with acronym in parentheses, irrespective of previous use
%%% \acs{label} acronym form, irrespective of previous use
%%% \acl{label} written out form without following acronym
%%% \acp{label} plural form of acronym by adding an s. \acfp. \acsp, \aclp work as well.
%%% \acs{label} Short: short version of the acronym
%%% \acfp, acsp aclp acfi acused acsu aclu
%%%%%%%%%%%%%%%%%%%%%%%%%%%%%%%%%%%%%%%%%%%%%%%%%%%%%%%%%%%%%%%%%%%

\begin{acronym}
	
\acro{5G}{fifth generation of mobile networks}
% A
\acro{ACRDA}{asynchronous contention resolution diversity ALOHA}%
\acro{APP}{a posteriori probability}%
\acro{ARA}{accumulate repeat accumulate}%
\acro{AWGN}{additive white Gaussian noise}%
% B

% C
\acro{CDF}{cumulative distribution function}
\acro{CN}{check node}
\acro{CRA}{contention resolution ALOHA}
\acro{CRC}{cyclic redundancy check}
\acro{CRDSA}{contention resolution diversity slotted ALOHA}
\acro{CSA}{coded slotted ALOHA}
% D
\acro{DAMA}{demand assigned multiple access}%
\acro{DSA}{diversity slotted ALOHA}%
% E
\acro{ECRA}{enhanced contention resolution ALOHA}%
\acro{eMBB}{enhanced mobile broadband}
\acro{EXIT}{extrinsic information transfer}
% F
\acro{FEC}{forward error correction}%
% G
\acro{GEO}{geostationary orbit}%
% H

% I
\acro{IC}{interference cancellation}%
\acro{IRCRA}{irregular repetition contention resolution ALOHA}%
\acro{IRSA}{irregular repetition slotted ALOHA} %
% K

% L
\acro{LDPC}{low-density parity-check}
\acro{LLR}{log-likelihood ratio}
\acro{LT}{Luby transform}
% M
\acro{M2M}{machine-to-machine}%
\acro{MAC}{medium access}%
\acro{MF}{matched filter}%
\acro{MF-TDMA}{multi-frequency time division multiple access}%
\acro{MRC}{maximal-ratio combining}
\acro{mMTC}{massive Machine Type Communications}
% N

% O

% P
\acro{PDF}{probability density function}%
\acro{PER}{packet error rate}%
\acro{PLR}{packet loss rate}%
\acro{PRACH}{physical random access channel}%
% Q
\acro{QPSK}{quadrature amplitude shift keying}
% R
\acro{RA}{random access}%
\acro{RV}{random variable}%
% S
\acro{SA}{slotted ALOHA}%
\acro{SB}{Shannon bound}%
\acro{SC}{selection combining}%
\acro{SIC}{successive interference cancellation}%
\acro{SINR}{signal-to-interference and noise ratio}%
\acro{SNIR}{signal-to-noise-plus-interference ratio}%
\acro{SNR}{signal-to-noise ratio}%
% T
\acro{TDMA}{time division multiple access}%
\acro{TS}{time sharing}%
% U

% V
\acro{VF}{virtual frame}
\acro{VN}{variable node}
% W
\acro{WER}{word error rate}
% X

\end{acronym}

%%%%%%%%%%%%%%%%%%%%%%%%%%%%%%%%%%%%%%%%%%
\section{Introduction}
\acresetall

Driven by the emerging \ac{M2M} communications and the Internet of things services, the number of connected devices is expected to reach the impressive number of $50$ billion by $2025$ \cite{5G_survey}. One of the key challenges \--- still largely unresolved in the current release of the 5G standard \cite{5g_release15} \--- is the problem on how to efficiently share the medium among a vast population of terminals intermittently and, possibly unpredictably, sending small data packets. The change in perspective required by the characteristic of \ac{M2M} data traffic calls for novel solutions to the medium access problem. The classic scheduled approach, well-suited for the transmission of large amounts of data, becomes rapidly inefficient due to the overhead required to assign resources to users. In particular, exchange of signaling information may become even larger than the data packet itself, and drastically lowers the efficiency of the medium access policy (see for example the \ac{PRACH} procedure \cite{Laya2014}).

A natural solution to reduce the signaling overhead, and thus increasing the achievable efficiency is to rely on random access techniques. In recent years, the classic ALOHA and \ac{SA} schemes \cite{Abramson1970,Roberts1975} have inspired a flourishing of novel medium access protocols that can be collectively labeled as \emph{modern random access} \cite{casini2007contention,Liva2011,Paolini2014,Sandgren2017,Stefanovic2013TCOM}. In all these schemes, nodes send proactively multiple copies of the same packet, while \ac{SIC} enhances the receiver and helps in resolving contention. The pioneer of these schemes is \ac{CRDSA} \cite{casini2007contention}\footnote{Concurrently also \cite{Giannakis_2007} proposed a random access tree-resolution scheme that relies on the use of \ac{SIC} at the receiver, which is able to largely outperform the classic standard and modified tree-algorithms \cite{Massey1981, Gallager1985}.} where the nodes are enforced to transmit $2$ packet copies per transmission attempt \cite{casini2007contention}. The \ac{CRDSA} protocol shows a great performance gain compared to \ac{SA}. Tools borrowed from the theory of codes on graphs \cite{Luby_2001, Richardson_2001} enable an asymptotic analysis of the performance \cite{Liva2011}, when the delay among replica grows very large. The analysis suggests that a variable number of packet copies per user is beneficial. The probability mass function of these packet copies  is called \emph{user degree distribution}. Analysis of the finite length performance (when the delay among replicas becomes bounded) have shown that good user degree distributions discovered for the asymptotic setting perform well also in the bounded delay regime \cite{Ivanonv_2015_Letter,i2018finite}. 

A further extension is proposed in \cite{Paolini2014}, where instead of simply \emph{repeating} the packets prior transmission, a coded version is generated. Such a modification is able to achieve higher energy efficiency, compared to \cite{Liva2011}. In \cite{Paolini2014}, also an achievable throughput region for the collision channel is derived. Impressively, letting the maximum number of replicas and delay among them grow large is sufficient to achieve the limit of $1$ packet per slot in the collision channel \cite{Narayanan2012}. This remarkable result shows that such schemes are able to close the gap to coordinated and orthogonal medium access schemes. Along a similar line of research, in \cite{Stefanovic2013TCOM} the authors investigate the behavior of a random access protocol with repetitions where the frame dimension is not set a-priori but it is dynamically adapted for maximizing the throughput.

In \cite{Polyanskiy2017} an information theoretic rate bound that considers the finite length nature of \ac{M2M} communications and a more realistic channel model is derived. The paper shed the light on the gap between practical schemes like \ac{CRDSA} and the achievable bound, raising awareness  to a larger community about the energy-efficiency problem. A new wave of research has been initiated, bringing several new approaches to the uncoordinated (and unsourced) multiple-access problem, e.g. \cite{Effros2018,Ordentlich2017,Vem2017,Calderbank2018,Amalladinne2018}.

The relaxation of  slot-synchronicity reduces the complexity and especially power consumption of the transmitter devices. Avoiding the need to keep synchronization to a common clock, enables longer \emph{sleep} times for the nodes. This is of outermost importance in many \ac{M2M} scenarios, where devices are powered via batteries that cannot (or may not) be replaced for their entire lifetime. Some recent solutions derived from the \emph{asynchronous} ALOHA protocol have shown to be competitive in comparison with their slot-synchronous counterpart \cite{Kissling2011a, DeGaudenzi2014, Clazzer18:ECRA, Duman19}. These solutions also adopt the transmission of multiple copies of the packets and \ac{SIC} at the receiver. When the receiver entangles combining techniques, e.g. maximal-ratio combining, with \ac{SIC} slot-synchronous protocols may be outperformed~\cite{Clazzer18:ECRA}.\footnote{The performance heavily depends on the specific configuration, i.e. number of replicas, physical layer forward error correction, channel conditions, etc.}

In this work, we consider an asynchronous random access scheme. We assume that every packet (codeword) is subject to \ac{AWGN} and possibly to interference. Due to the asynchronous nature of the scheme the interference may affect different portions of the packet with different  power, depending on the number of colliding users. In order to allow reliable transmission on the asynchronous random access channel, a suitable error protection scheme needs to be used. Works in the literature usually assume capacity achieving random code ensembles and apply a threshold based model for decoding \cite{Kissling2011a}: the average signal to noise plus interference ratio over a packet is compared to the Shannon limit, i.e., the worst channel parameter for which error free transmission is possible, to decide whether decoding is possible or not.  In \cite{KC13}, the authors replace the Shannon limit by iterative decoding thresholds of some  off-the-shelf \ac{LDPC} code ensembles. In an earlier conference paper \cite{MCJ+18}, we considered a decoding region (which can be seen  as a  multi-dimensional threshold based model) to perform dedicated protograph-based \ac{LDPC} code design for the asynchronous random access channel and to estimate the \ac{PLR} of the random access protocol. However, it remains an open question how accurately such threshold based models can predict the \ac{PLR}.

The code design of \cite{MCJ+18} using a multi-dimensional threshold based model can be seen as a type of code design for unequal error protection. Unequal error protection using error correcting codes has long been studied in the literature for different channels. Early works date back to the 1960s. For instance, \cite{Wolf1967} addresses the problem of achieving different bit error probabilities in different parts of the decoded codeword.  More recently, both turbo codes and \ac{LDPC} codes have been designed for channels that introduce different reliabilities on different parts of the codeword \cite{Caire1998, Boutros2005, Boutros2010} to counteract the effect of block fading:  a block is characterized by a constant fading coefficient, while block by block the fading coefficient is independently drawn from a predefined distribution.

In the following,  we focus on the asynchronous random access channel for which we provide tailored protograph \ac{LDPC} code designs. Here, we make use of a  surrogate channel model to simplify the code design. We  show that the resulting codes evidently boost the \ac{MAC} performance, even in comparison with the most competitive code designs for standard \ac{AWGN} channels. As an extension of \cite{MCJ+18}, we do not  restrict to finding \ac{LDPC} code ensembles with favorable iterative decoding thresholds only, but also design finite-length codes. These codes are used to simulate the physical layer of the random access protocol, giving a more realistic estimate of the \ac{PLR}. 
The key contributions of the paper can be summarized as follows:
\begin{itemize}
    \item In Section~\ref{sec:code_ds} we present a surrogate channel model, exploited in the code design phase, which assumes constant interference power over a fraction $\alpha$ of the codeword. To facilitate code design, we further approximate the aggregate interference contribution, possibly generated by a multitude of terminals, as Gaussian. In Section~\ref{sec:code_design}, we present a protograph \ac{LDPC} code design for this channel model and its iterative decoding threshold is compared with the one of a raptor-like \ac{LDPC} code design proposed for the recently introduced \ac{5G} standard. Both code designs are compared with the Shannon limit.
    \item The impact of the Gaussian interference assumption on the code performance is also considered in Section~\ref{sec:IM_code_design}. The expression of the \ac{LLR} and the threshold performance, computed with quantized density evolution, are presented for a single interferer when the Gaussian assumption is dropped. 
    \item In order to get a first \--- yet not fully accurate \--- performance characteristic for the proposed LDPC codes in the \ac{RA} channel, we elaborate on the decoding condition so as to abstract the physical layer in Section~\ref{sec:apl}. A decoding region, as a function of the interference pattern for both a random code ensemble and for the \ac{LDPC} code ensemble is derived. Although more accurate than the surrogate channel model, since the effective interference power and affected codeword position is considered, the abstraction grounds on the iterative decoding threshold, and thus on large blocks assumption. 
    \item Since \ac{RA} is particularly appealing for short packet transmission, we depart from the physical layer abstraction, and present in Section~\ref{sec:results_phy} physical layer simulations considering finite block length. Interestingly, the codes designed for the surrogate channel model still perform very well on the asynchronous multiple access channel. Moreover, the performance trends and relative performance identified via the simpler simulations with the abstracted physical layer are confirmed.
\end{itemize}

\section{System Model}
\label{sec:system_model}

In this Section we describe the system model, starting with the medium access policy in \ref{sec:CRA}. Section \ref{sec:plm} follows with the physical layer model.

\subsection{Asynchronous Random Access Protocol}
\label{sec:CRA}

We consider an uncoordinated asynchronous random access protocol based on \cite{Kissling2011a,DeGaudenzi2014}.\footnote{With respect to \cite{Kissling2011a}, no concept of \ac{MAC} frame is present, so that the terminals operate in a full asynchronous scenario. Compared to \cite{DeGaudenzi2014}, local time slots constraining the transmission delay between replicas of the same user are eliminated.} An infinite user population generates data traffic modeled as a Poisson process with intensity $\load$, called the \emph{channel load}. It is measured in packet arrivals per packet duration $\pkLen$. Specifically, the probability distribution that $u$ users initiate a transmission within a packet duration is $P(u)=\frac{\load^u e^{-\load}}{u!}$.

Users are allowed to make a single attempt to transmit their data, i.e., no re-transmissions are considered. Prior to transmission, the data packet is replicated $\dg=2$ times. We refer to each copy as a \emph{replica}. The replicas are transmitted adhering the following rules:
\begin{enumerate}
    \item self-interference must be avoided, i.e. no portion of the two replicas shall overlap;
    \item the maximum delay \--- called \ac{VF} \--- between the start of the first replica and the end of the second one shall not exceed $\vf$ seconds;
    \item the delay between the start of the two replicas is drawn uniformly at random within the interval $\left(t_0^{(u)}+\pkLen,\, t_0^{(u)}+\vf-\pkLen\right]$, with $t_0^{(u)}$ the activation time of user $u$.
\end{enumerate}
Once the transmission times for the replicas are chosen, this information is stored in the header in order to enable \ac{SIC} at the receiver. Then the replicas are encoded so to be protected against noise and multiple access interference.

\begin{figure}
\centering
\includegraphics[width=\columnwidth]{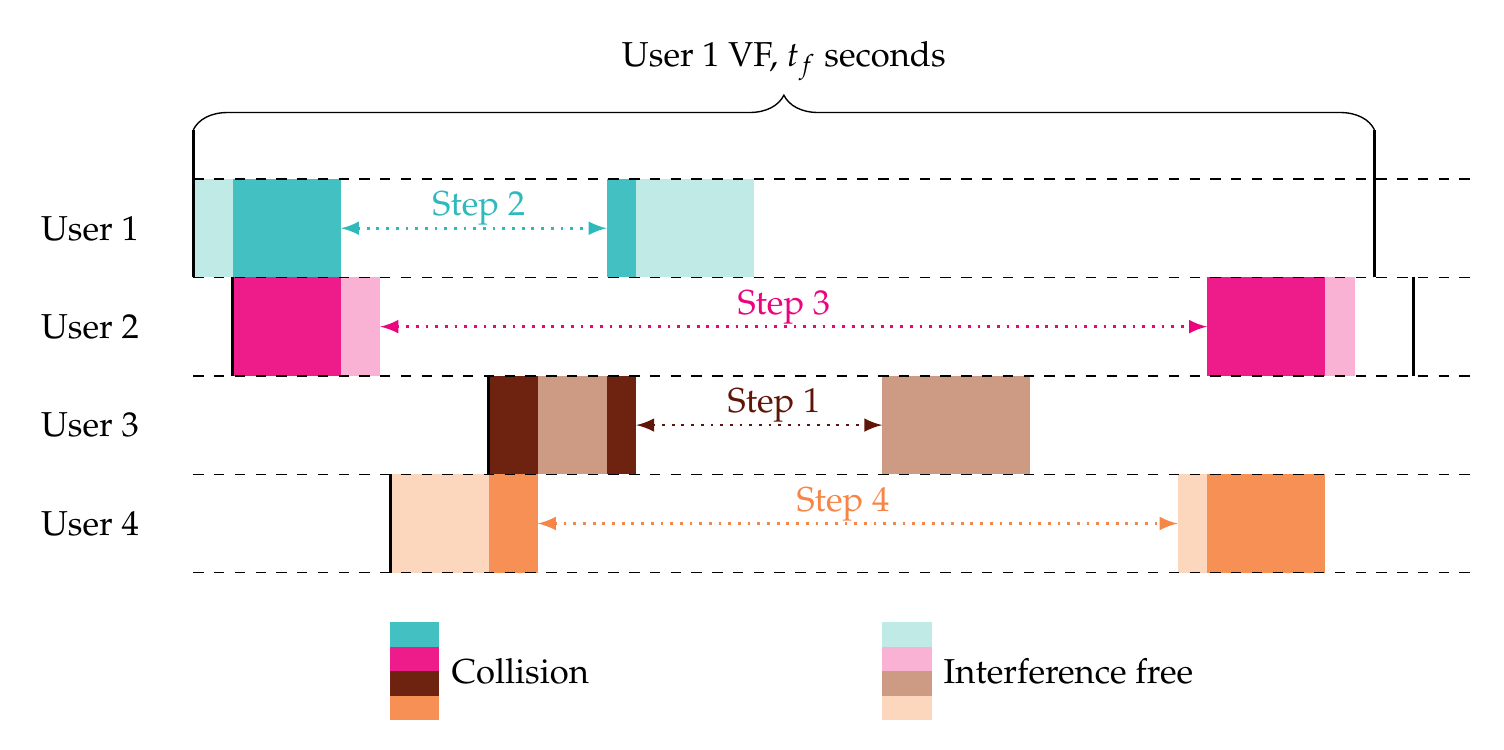}
\caption{Example of collision pattern at the receiver, and of the \ac{SIC} procedure. Upon correct decoding, \ac{SIC} removes the packet and its twin.}
\label{fig:MAC_frame}
\end{figure}

At the receiver side, the incoming signal is sampled, stored and subsequently processed. The receiver makes use of a decoding window of size $\wind$ samples. At first, replicas are detected, e.g., with correlation-based rules \cite{Chiani_2010,Clazzer2017_Corr}.\footnote{In this work ideal detection is considered, i.e., all transmitted replicas can be detected by the receiver and no false alarms are present.} Channel estimation is performed (for details see Section \ref{sec:plm}). The soft-demodulator provides  bit-wise \acp{LLR} as input to the channel decoder. When the \ac{CRC} matches, the replica is declared as correctly decoded. When this is the case, readily two operations follow:
\begin{enumerate}
    \item the replica waveform is reconstructed on a sample level and subtracted from the incoming signal;\footnote{We will consider ideal interference cancellation, i.e. after cancellation no residual power is left by the replica.}
    \item the information about the twin location (the time position of the other replica of the same user) is extracted from the header.
\end{enumerate}
The second operation is followed by data-aided channel estimation on the packet copy. In fact, the entire data carried by the packet is now known and can be used as pilots for refining channel estimation. The received waveform of the twin can be reconstructed and removed from the received signal. We shall note that if a replica can be decoded, the interference reduction triggered by \ac{SIC} may impact underlying collided packets and can lead to further packet recovery. The receiver proceeds extracting the second candidate replica and repeats the aforementioned operations. \ac{SIC} is iterated until no more packets are present in the decoding window, or when a predefined maximum number of iterations is reached. The second option is normally adopted when the receiver has tighter complexity or delay constraints. Once the operations on the current receiver decoding window are terminated, the receiver window is shifted forward by $\Dwind \ll \wind$ samples, and the detection of the replicas is initiated again.

 \begin{Example}
Consider an instance of \ac{SIC} as depicted in Figure~\ref{fig:MAC_frame}. A \ac{SIC} step consists of  successfully decoding  a replica and removing it as well as its twin from the received signal. Thus, in Figure~\ref{fig:MAC_frame}, the first \ac{SIC} step consists of decoding the second replica of user $3$, since it is interference free. Then, the contribution of both replicas can be removed from the received signal. Consequently, the second replica of user $1$ and the first replica of user $4$ become interference-free. We decode user $1$, remove both replicas from the signal. We proceed in the same way for user $2$. Since no interference is present anymore, we assume that also the user $4$ can be successfully decoded.
\end{Example}

Note that replicas with the lowest level of interference are the ones with the highest chance to be correctly decoded. Interference-free replicas or replicas collided with a single interferer have the highest chance to be recovered in a \ac{SIC} iteration. Considering the latter, collision of \emph{two} packets yields interference which is either at the beginning or at the end of a packet. Consequently, an error-correcting code able to better protect the beginning or the end of a packet, should come to the aid of \ac{SIC} improving the random access overall performance. In light of this, in Section~\ref{sec:code_ds} we focus on the design of error correcting codes able to better protect the beginning and end of a codeword.

\subsection{Asynchronous Random Access Channel Model}
\label{sec:plm}
\begin{sloppypar}
In the following, we define the general channel model for the considered random access scheme. Assume that all replicas of all users form the set $\rSet$. Consider the $\rep$-th replica in $\rSet$. The transmitted signal $\tx^{(\rep)}$ corresponding to the modulated codeword ${\bm{\symVal}^{(\rep)}=\left( \symVal_0^{(\rep)}, \symVal_1^{(\rep)}, \dots, \symVal_{\ns-1}^{(\rep)} \right)}$ is
\end{sloppypar}
\begin{equation}
\label{eq:tx_sig}
\tx^{(\rep)}(\tm)= \sum_{i=0}^{\ns-1} \symVal_i^{(\rep)} \pulse(\tm-i \sLen)
\end{equation}
where $\ns$ is the number of modulation symbols in a packet, $\sLen$ is the symbol duration, and $\pulse(\tm)$ is the pulse shape. The received signal is in general affected by a frequency offset, modeled as a uniformly distributed random variable $\freq^{(\rep)} \sim \mathcal{U}\left[-f_{\mathrm{m}};f_{\mathrm{m}}\right]$, with $f_{\mathrm{m}}$ the maximum frequency offset. A sampling epoch, also modeled as a uniformly distributed random variable $\epoch^{(\rep)} \sim \mathcal{U}\left[0;\sLen \right)$ \cite[Ch. $2$]{Mengali_Dandrea}. Both frequency offset and sampling epoch are common to each replica of the same user, but independent user by user. Furthermore, the signal is also affected by phase offset $\phase^{(\rep)}$, modeled as a uniformly distributed random variable  between $0$ and $2\pi$, i.e., $\phase^{(\rep)} \sim \mathcal{U}\left[0;2\pi\right)$. The phase offset is assumed to be independent replica by replica. Further, we assume \ac{AWGN}. No fading is considered which is typical, e.g., for fixed-terminal geostationary orbit satellite scenarios. For $f_{\mathrm{m}}\sLen \ll 1$, the received signal $\rx(\tm)$ after matched filtering, which is a superposition of all replicas $\rep \in \rSet$, is
\begin{equation}
\label{eq:rx_sig}
\rx(\tm) = \sum_{\rep \in \rSet} \tilde{\tx}^{(\rep)}(\tm - \epoch^{(\rep)} - t_0^{(\rep)}) e^{j\left(2\pi \freq^{(\rep)} + \phase^{(\rep)}\right)} + \noisev(\tm)
\end{equation}
where  $\tilde{\tx}^{(\rep)}(\tm)$ is the matched filtered signal $\tx^{(\rep)}(\tm)$. In equation~\eqref{eq:rx_sig}, $t_0^{(\rep)}$ is the $\rep$-th replica delay w.r.t. the common reference time. The noise term $\noisev(\tm)$ is given by $\noisev(\tm) \triangleq \noisePr(\tm) \ast \pulse(\tm)$, where $\noisePr(\tm)$ is a white Gaussian process with single-sided power spectral density $\noiseSD$. For the sake of simplicity, throughout the paper we make the assumptions that $\freq^{(\rep)}=0$ and $\epoch^{(\rep)}=0$, $\forall \rep \in \rSet$. Hence, equation~\eqref{eq:rx_sig} becomes
\begin{equation}
\label{eq:rx_sig2}
\rx(\tm) = \sum_{\rep \in \rSet} \tilde{\tx}^{(\rep)}(\tm - t_0^{(\rep)}) e^{j\phase^{(\rep)}} + \noisev(\tm).
\end{equation}

We focus again on replica $\rep$. After ideal detection, the received signal corresponding to $\rep$ can be identified and isolated. Ideal channel estimation recovers perfectly the phase offset $\phase^{(\rep)}$. After compensation, the discrete-time version of the received signal  ${\rxVec^{(\rep)} = (\rx_0^{(\rep)}, \dots, \rx_{\ns-1}^{(\rep)})}$ corresponding to the replica $\rep$ is given by
\begin{equation} \label{eq:rec_y}
\rxVec^{(\rep)} = \bm{\symVal}^{(\rep)} + \intVec^{(\rep)} + \noiseVec^{(\rep)}.
\end{equation}
Here $\noiseVec^{(\rep)}=(\noisev_0^{(\rep)},\dots,\noisev_{\ns-1}^{(\rep)})$ are the samples of a complex discrete white Gaussian process with ${\noisev_i^{(\rep)} \sim \mathcal{CN}(0,2\noise)}$, $\forall i\in \{0,\dots,\ns-1\}$, $\forall r\in \rSet$. The aggregate interference contribution over the replica-$\rep$ signal is $\intVec^{(\rep)}=\left(\intVal_0^{(\rep)},\dots,\intVal_{\ns-1}^{(\rep)}\right)$, with
\begin{equation}
\label{eq:int_aggr_samp}
\intVal_i^{(\rep)} = \sum_{\repInd \in \rSet \setminus \rep} \tilde{\tx}^{(\repInd)}(k\sLen - \Delta t_0^{(\repInd)}) e^{j\Dphase^{(\repInd)}}\qquad\text{for all } i\in \{0,\dots,\ns-1\}.
\end{equation}
Here, $ \Delta t_0^{(\repInd)}=t_0^{(\rep)}-t_0^{(\repInd)}$ and $\Dphase^{(\repInd)}=\phase^{(\rep)}-\phase^{(\repInd)}$. The instantaneous received power for symbol $i$ is $\sPw_i^{(\rep)}\triangleq\mathbb{E}\left[|\symVal_i^{(\rep)}|^2\right]$. The useful received power is assumed to be constant over the entire replica $\rep$, i.e. $\sPw_i^{(\rep)}=\sPw^{(\rep)}$ for $i=0,\dots,\ns$ and for $\forall \rep \in \rSet$. Users are received with the same power thanks to perfect power control, i.e. $\sPw^{(\rep)}=\sPw$ for $\forall \rep \in \rSet$. As a result $\sPw_i^{(\rep)}=\sPw$. The aggregate interference power for symbol $i$ is $\iPw_i^{(\rep)}\triangleq\mathbb{E}\left[|\intVal_i^{(\rep)}|^2\right]$. Finally, the noise power is $2 \noise$ and we define the signal-to-noise ratio, $\snr = \sPw/ 2 \noise$. We also define the signal-to-interference ratio $\gamma_i=\sPw/\iPw_i$. 					
\section{Code Design for the Asynchronous Random Access Channel}
\label{sec:code_ds}
\subsection{Protograph LDPC Codes}
We propose protograph-based binary \ac{LDPC} code designs tailored to the interference that occurs in an asynchronous \ac{RA} scenario. The parity-check matrix of these codes is derived from a relatively small matrix, called a base matrix  $\basem{}$, which represents the code constraints. Alternatively, we may describe the base matrix as a bipartite graph, termed protograph, with $n_b$ \acp{VN} and $m_b$ \acp{CN}. Non-zero entries $b_{i,j}$ in $\basem{}$ represent connections between \acp{VN} of type $j$ and \acp{CN} of type $i$. In order to improve the code performance, we may decide to puncture some of the \ac{VN} types in the base matrix whose number is denoted by $p_b$. The number of not punctured variable nodes is denoted by $(n_b-p_b)$. The parity check-matrix of a protograph based $(n,k)$ \ac{LDPC} code is obtained by expansion or lifting of the base matrix. This is done by copying the protograph $n/(n_b-p_b)$ times and interconnecting the copies among each other following certain rules (see \cite{Tho03} for details). In this work, we consider standard Gray labelled \ac{QPSK} modulation where the binary labels of a symbol consist of two bits. Thus a packet with $\ns$ symbols is protected by an LDPC code with codeword length $n=2\ns$ bits.

\subsection{Code Optimization}
Good base matrices are found by some optimization algorithm, such as differential evolution \cite{Uch14}. The goal is to find protographs which maximize a gain function. To this end, differential evolution creates a generation of candidate base matrices for which the gain function is evaluated. By introducing perturbations on the base matrix entries, a new generation of base matrices is obtained. They may replace the original ones if their gain function is higher. After a certain number of generations we choose those base matrices which maximize the gain function.

The gain function is usually a function of the protograph ensemble's iterative decoding threshold which can be seen as the worst channel parameter for which symbol error probability vanishes (assuming that the block length and the number of decoding iterations go to infinity). We will give a more precise description in the following.

Iterative decoding thresholds can be found using (quantized) density evolution \cite{Richardson_2001}, or a suitable approximation, such as \ac{EXIT} analysis \cite{LC07} if certain Gaussian assumptions can be made. \ac{EXIT} analysis simplifies the computation of the protograph code ensemble's compared to (quantized) differential evolution. To compute iterative decoding thresholds, we make use of the channel \acp{LLR} distributions for each codeword bit. In the following sections, we discuss the effect of the interference model on the channel \ac{LLR} distributions and thus on the code design.

\subsection{Simplified Channel Models for Code Design} \label{sec:surrogate}
\begin{figure}
\centering
\includegraphics[width=0.9\columnwidth]{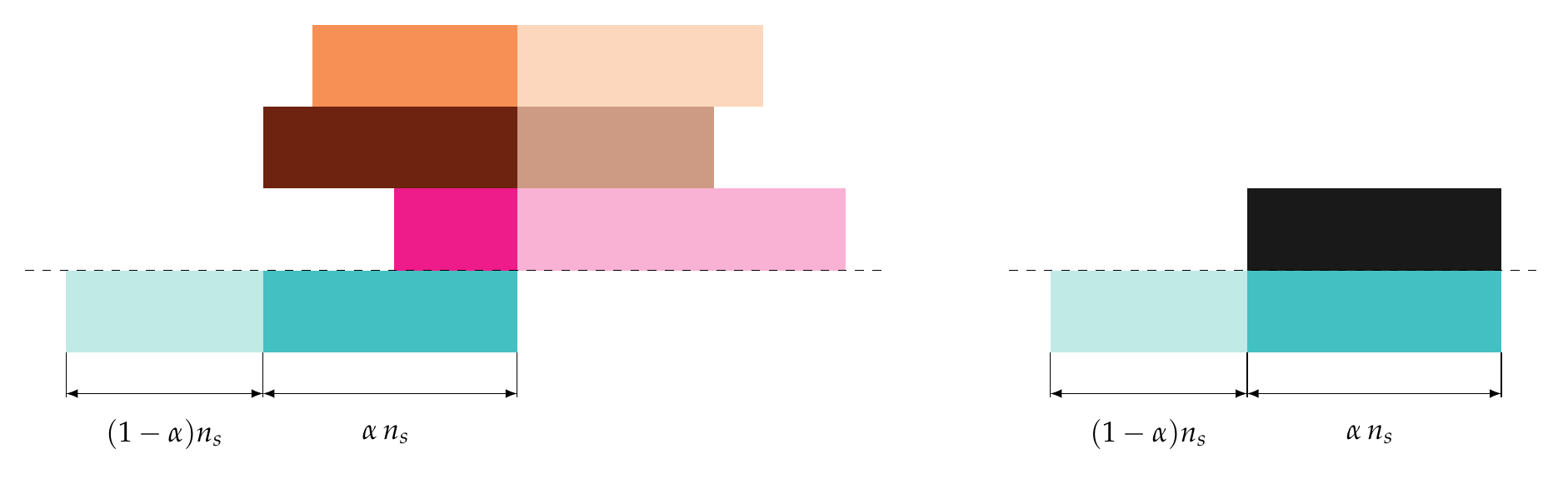}
\caption{Collision of multiple users (left) and abstracted surrogate model with constant interference power over $\alpha \ns$ symbols (right).}
\label{fig:collision}
\end{figure}

To facilitate the code design, we assume the following \emph{surrogate} channel model:
\begin{itemize}
    \item A fraction $1-\alpha$, $0\leq \alpha \leq 1$, of a (modulated) codeword  is only affected by noise with power $2  \noise$.
    \item A fraction $\alpha$ of a (modulated) codeword is affected by noise and interference of constant power over the fraction.
\end{itemize}
Thus, we make the assumption of a \emph{block interference channel} \cite{McEliece1984} which is a widely used model in the literature (see, e.g., \cite{Clazzer18:ECRA} and references therein). Clearly, this is a simplification, since  a codeword in reality may experience various interference levels due to multiple packet collisions. Note also that for our asynchronous \ac{RA} protocol, the interference is confined to the beginning, the end, or to the beginning and the end of a codeword. This observation is exploited for the code design.  The surrogate channel model is exemplified in Figure~\ref{fig:collision}.  We will illustrate its validity Section~\ref{sec:results}.

\subsubsection{Gaussian Interference Model}
\label{sec:gim}
Consider the interference vector $\intVec^{(\rep)}$ in \eqref{eq:rec_y}. Let its $i$-th element $\intVal_i^{(\rep)}$ be non-zero, i.e., the $i$-th packet symbol is subject to interference. Then, $z^{(\rep)}_i$ is a sample of a complex Gaussian distribution $\mathcal{CN}(0,2\inter)$, where $2 \inter$ is the interference power. This assumption is accurate already in case of a few interferers having different phase, frequency and time offset. When only one interferer affects a packet, the assumption becomes inaccurate. Nevertheless, we will illustrate that it is a reasonable model for the \ac{LDPC} code design.

Under the assumption of  Gaussian interference, the interference plus noise also follows a complex Gaussian distribution $\mathcal{CN}(0,2\nPI)$, with
\begin{equation}
\label{eq:noise_inter}
\nPI=\noise+\inter.
\end{equation}
Let us define as $\Ca\left(\nPI\right)$ the \ac{QPSK} constraint \ac{AWGN} channel capacity for a given $\nPI$. The corresponding outage capacity $\Capa$ for our surrogate channel model in Section~\ref{sec:surrogate} under the Gaussian interference assumption becomes
\begin{equation}
\label{eq:TSC}
\Capa(\alpha, \noise, \nPI) = (1-\alpha) \,\Ca\left(\frac{1}{2\noise}\right) + \alpha \, \Ca\left(\frac{1}{2\nPI}\right).
\end{equation}
For a fixed $\alpha$ and $\noise$ we can invert \eqref{eq:TSC} and write $ \nPI$ (or likewise $\inter$) as a function of  $\Capa$.  Then, for a fixed rate, error free transmission is possible if the interference power $\inter<\interCap$, where $\interCap$ is referred to as the Shannon limit.

Code design for the case of the Gaussian interference model becomes equivalent to code design for an \ac{AWGN} channel where codeword bits are subjected to one of two possible noise levels. The design of \ac{LDPC} photograph codes for this channel is well studied and, given the interference is modeled as Gaussian noise, various approximations to density evolution, such as \ac{EXIT} charts, are well verified in this scenario.

\subsubsection{Single Interferer Model} \label{sec:non_Gaussian_interferer_model}

The Gaussian interference model is a reasonable approximation already in presence of a few interferers under the assumption of time, frequency and phase offset. When a single interferer is present over the codeword, the Gaussian model may become imprecise and the interference shall be better characterized, taking into account the chosen modulation. Thus in this Section, we consider a codeword affected by interference generated by a single \ac{QPSK}-modulated interferer, received with the same power. After ideal detection and ideal channel estimation, the interference occuring on a codeword symbol $i$ presents a relative phase shift $\Dphase^{(1)}$, and a relative epoch $0<\Depoch^{(1)}<\sLen$. Hence, the interference on codeword symbol $i$, i.e. $\intVal_i$, can be expressed as
\begin{equation}
\label{eq:int_samp_one_int}
\intVal_i = \tilde{\tx}^{(1)}(k\sLen - \Depoch^{(1)} - \Delta t_0^{(1)}) e^{j\Dphase^{(1)}}.
\end{equation}
Note that $\gamma_i=\sPw/\iPw_i=1$. For ease of notation we drop the superscript $^{(1)}$, so that $\Dphase^{(1)}\triangleq\Dphase$ and $\Depoch^{(1)}\triangleq\Depoch$.
\begin{figure}[!t]
	\centerline{
		\includegraphics[width=0.8\linewidth]{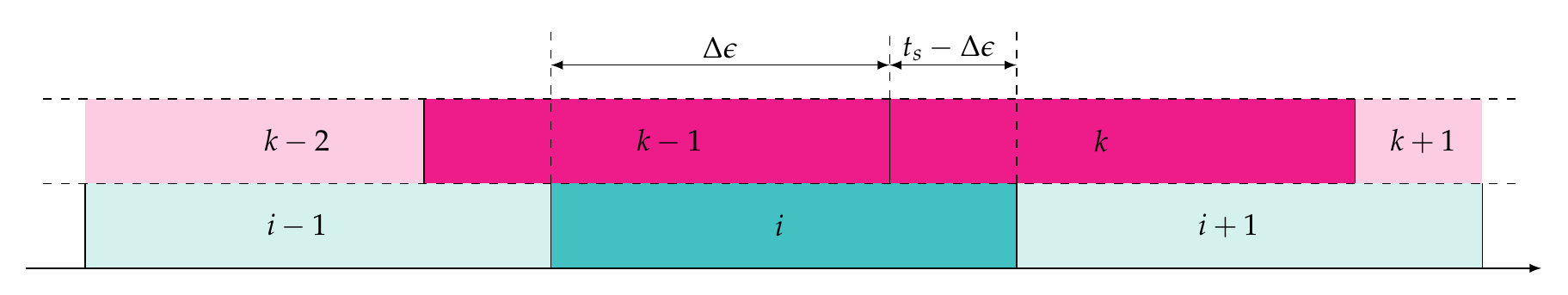}}
			\caption{The $i$-th codeword symbol (azure) affected by a single interferer (red), with a relative epoch of $\Depoch$. Both interferer symbol $k-1$ and $k$ impact the codeword symbol $i$.}
	\label{fig:symbol_mis}
\end{figure}
Figure~\ref{fig:symbol_mis} shows the considered scenario. The $i$-th codeword symbol is interfered by a portion $\Depoch$ of symbol duration with the $(k-1)$-th symbol of the interferer and by $\sLen-\Depoch$ with the $k$-th symbol of the interferer. We define $\mathcal{S} = \{S_1,S_2,S_3,S_4\}$ as the set of \ac{QPSK} constellation points with respective labels  $\{00,01,11,10\}$. For a given relative epoch $\Depoch$, and a relative phase offset $\Dphase$, known at the receiver, the channel \ac{LLR} of bit level 1 of a Gray labeled QPSK signal are given by
\begin{align}
&L_1(y,\Dphase,\Depoch')=\\
&\log\left[\frac{\displaystyle{\sum_{\{S_{k-1},S_{k}\}\in\mathcal{S}^2}}\textstyle \left(e^{-\frac{\left|y-S_1- e^{j\Dphase}\left(\Depoch' S_{k-1}+(1-\Depoch')S_{k}\right)\right|^2}{2 \noise}}+ e^{-\frac{\left|y-S_2- e^{j\Dphase}\left(\Depoch' S_{k-1}+(1-\Depoch')S_{k}\right)\right|^2}{2 \noise}}\right)}{\displaystyle{\sum_{\{S_{k-1},S_{k}\}\in\mathcal{S}^2}}\textstyle\left(e^{-\frac{\left|y-S_3- e^{j\Dphase}\left(\Depoch' S_{k-1}+(1-\Depoch')S_{k}\right)\right|^2}{2 \noise}}+e^{-\frac{\left|y-S_4- e^{j\Dphase}\left(\Depoch' S_{k-1}+(1-\Depoch')S_{k}\right)\right|^2}{2 \noise}}\right)}\right]
\label{eq:llr_qpsk}
\end{align}
Where we defined $\Depoch'\triangleq \Depoch/\sLen$. For the sake of simplification, we assume symbol-synchronous interference i.e. $\Depoch=\Depoch'=0$ and unknown, uniformly distributed relative phase offset. We can write the the channel \ac{LLR} of bit level 1 as:

\begin{align}
L_1(y,\Depoch=0)=\log\left[\frac{\displaystyle{\sum_{S_{k}\in\mathcal{S}}}\int_{-\pi}^{\pi}\textstyle\left(e^{-\frac{\left|y-S_1 -S_k e^{j\Dphase}\right|^2}{2 \noise}}+e^{-\frac{\left|y-S_2 -S_k e^{j\Dphase}\right|^2}{2 \noise}}\right)d\theta}{\displaystyle{\sum_{S_{k}\in\mathcal{S}}}\int_{-\pi}^{\pi}\textstyle\left(e^{-\frac{\left|y-S_3 -S_k e^{j\Dphase}\right|^2}{2 \noise}}+e^{-\frac{\left|y-S_4 -S_k e^{j\Dphase}\right|^2}{2 \noise}}\right)d\theta}\right].
\label{eq:llr_qpsk_rand}
\end{align}

\begin{figure}[!t]
\begin{center}
\begin{subfigure}{0.48\textwidth}
\centering\captionsetup{width=.9\columnwidth}
\includegraphics[width=\columnwidth]{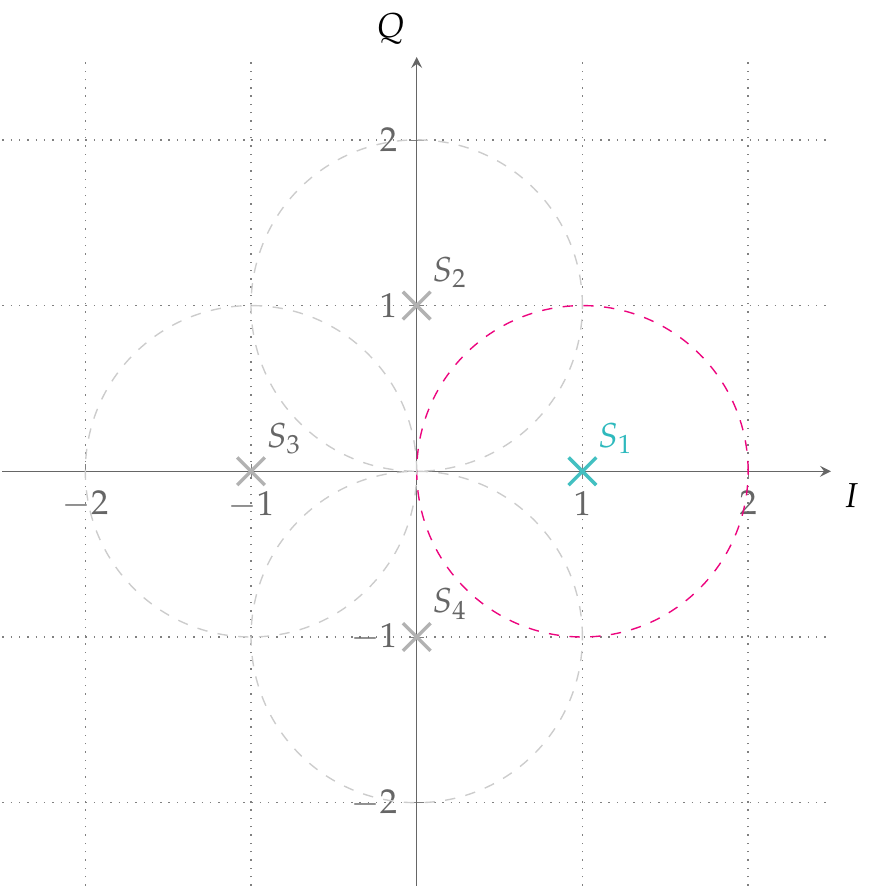}
   \caption{\ac{QPSK} modulated constellation after phase compensation at the receiver side. We highlight symbol $S_1=(1,0)$. If the symbol is affected by \ac{QPSK} modulated interference with the same unit power, no noise, and random phase the received symbol will lay on the red unit circle centered in $(1,0)$.}
   \label{fig:vul_per1}
\end{subfigure}
\begin{subfigure}{0.48\textwidth}
\centering\captionsetup{width=.9\columnwidth}
\includegraphics[width=\columnwidth]{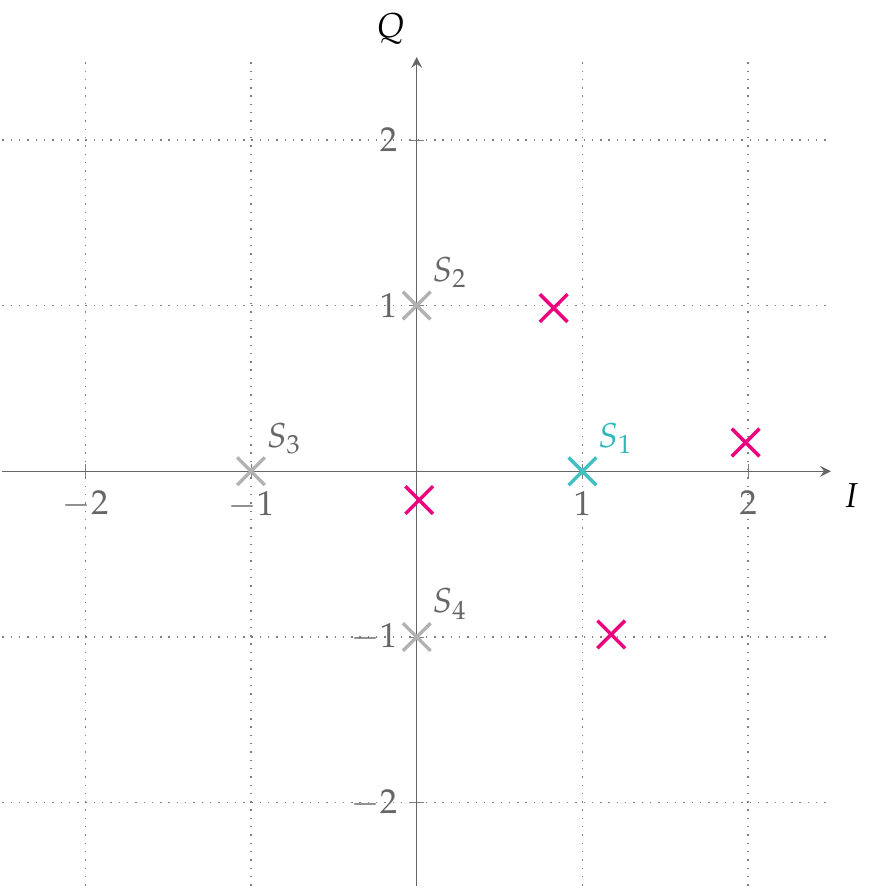}
   \caption{\ac{QPSK} modulated constellation after phase compensation at the receiver side and one \ac{QPSK} modulated interferer with the same unit power, no noise, and $\Dphase=10^\circ$ relative phase shift. Red crosses represent the four possible received symbols if the considered transmitted symbol was $S_1=(1,0)$.}
   \label{fig:vul_per2}
\end{subfigure}
\end{center}
\caption{\ac{QPSK} modulated constellation after phase compensation at the receiver side and single symbol-synchronous interferer with same power, relative phase shift and no noise.}
\label{fig:qpsk}
\end{figure}

Figure~\ref{fig:qpsk} shows the possible received symbol $i$ after phase compensation with no noise when symbol $S_1$ (or $S_2,S_3,S_4$) is transmitted. The received symbol can lay anywhere on the red circle if no assumption on interference relative phase shift is done, see Figure~\ref{fig:vul_per1}. If a relative shift of $\Dphase=10^{\circ}$ is present, then any of the four red points can be received depending on the transmitted interference symbol $k$, see Figure~\ref{fig:vul_per2}.

\begin{figure}[!t]
	\centering
	\includegraphics[width=.9\columnwidth]{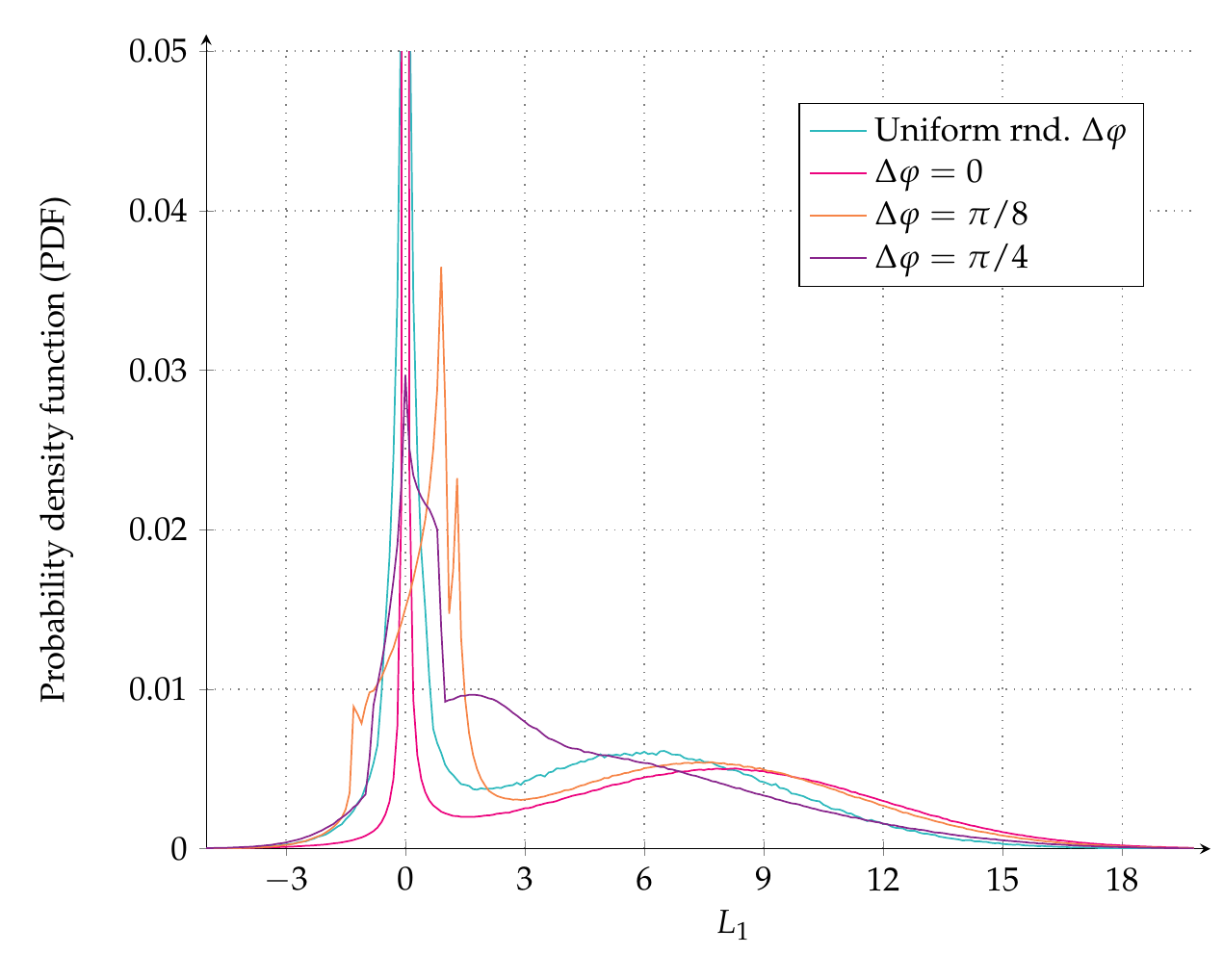}
	\caption{Probability density function of the bit \acp{LLR} assuming AWGN and symbol-synchronous QPSK interferer with unifrom at random phase (or with a fixed phase offset) when $\snr=6$~dB.}
	\label{fig: QPSK PDF}
\end{figure}

In Figure~\ref{fig: QPSK PDF} the \ac{PDF} of the bit \acp{LLR} assuming a symbol-synchronous \ac{QPSK} interferer with either uniform at random phase or with a fixed relative phase offset plus \ac{AWGN} noise with $\snr=6$~dB is shown. For the plot we assume that either symbol $S_0$ or $S_1$ has been transmitted, i.e., bit level one is zero. The curve with $\Dphase=0$ shows the \ac{PDF} of the bit \acp{LLR} for the case of aligned phase. The channel \ac{LLR}s are no-longer Gaussian and thus quantized density evolution, which passes the entire LLR though the decoding algorithm, rather than a Gaussian model, will be used to calculate the code thresholds.

\subsection{Code Design for Gaussian Interference} \label{sec:code_design}

Two different code designs for the surrogate channel with Gaussian interference, as defined in Section~\ref{sec:gim}, are presented next. The designs are exemplified for a code rate of $R_c=1/2$, \ac{QPSK} modulation and a target $\snr=6$~dB, typical parameters for an asynchronous \ac{RA} scheme \cite{Kissling2011a, DeGaudenzi2014}. However, the code design can be easily performed for any other parameter set.

 \subsubsection{Ad-hoc \ac{LDPC} Code Design}

 Observe that the interference present in the random access channel may affect the beginning, the end, and sometimes both beginning and end of a packet. In fact, by analyzing the interference patterns, we found that the third case, where the interference affects both the beginning and end of a packet, is a case of limited interest for the code design (although it is not difficult to embed additional constraints in the code design). To see this, consider the collision of two packets. Clearly, the interference here hits the beginning or end of a packet. Consider now that three packets collide such that there exists a packet with interference both at the beginning and at the end. Then, the two remaining packets will experience a lower level of interference which is either at the beginning or at the end. So \ac{SIC} may preferably decode one of those packets, and cancel their contribution from the received signal. Similar considerations can be made when four packets collide and we have a packet with both interference at the beginning and the end. For five or more interfering packets there could conceivably be cases where a packet with interference at both the beginning and end might be favorable to decode first, but for the code design we focus on the most frequent cases. Thus, we tailor our code design to these specific interference patterns by targeting codes which are robust w.r.t.\ interference at the beginning or at the end of a codeword.

 In the following, we target code designs robust to interference at the beginning or end of a codeword by imposing certain symmetry constraints on the code's base matrix $\basem{A}$.  Let us split $\basem{A}=[\basem{P}|\basem{Tx}]$ into two submatrices, an $m_b \times p_b$ matrix $\basem{P}$ for the punctured columns and an $m_b \times (n_b-p_b)$ matrix $\basem{Tx}$ composed by all columns which are not punctured. For the entries $b_{i,j}$ of $\basem{Tx}$ we impose
  \begin{equation} \label{eq:symmetry}
 b_{(m_b-i-1),(n_b-p_b-j-1)}\stackrel{!}{=}b_{i,j}
 \end{equation}
 $\forall i \in \{0,\ldots,m_b-1\}$ and  $\forall j \in \{0,\ldots,n_b-p_b-1\}$. The symmetry requirement in \eqref{eq:symmetry} states that the $j$-th column  of $\basem{Tx}$ from the left shall be equal to the $j$-th column from the right, whose elements are however placed in a reversed order. A similar requirement is put on the submatrix $\basem{P}$.

 For the protograph search by means of differential evolution we fix $n_b$, $m_b$, and $p_b$, such that the code rate $R_c =\frac{n_b-m_b}{n_b-p_b}$. We also fix $\noise$ (e.g., as a result of link budget considerations) and look for protographs which for certain $\alpha$, allow successful decoding at an interference power $2\inter$ as high as possible. Thus, the iterative decoding threshold is the maximum interference power for a certain $\alpha$ and $\noise$, such that the probability of symbol error in a codeword vanishes. Since $\alpha$ is the outcome of random process, we are interested in a code design that is robust for various values of $\alpha$, denoted by $\alphal$. This implies a multi-target optimization.  Due to complexity reasons, we resort to only a few different $\alphal$ for which we simultaneously optimize our base matrices.

 The Gaussian interference model yields channel \acp{LLR} for the two \ac{QPSK} bit-levels which are Gaussian distributed. For this setup,  \ac{EXIT} analysis \cite{LC07}  is known to provide accurate iterative decoding threshold estimates. Thus, for each  { $\alphal$} we aim at determining the maximum amount of interference ${\idt(\alphal)}$, such that error-free decoding is possible. Following the results in $\cite{LC07}$, this requires that the log-likelihood \ac{APP} mutual information ${\IAPP (j) \rightarrow 1}$ $\forall j \in \{0,\ldots, n_b-1 \}$ (assuming that the block length and the number of decoding iterations go to infinity).

 {For a general code ensemble, the computation of the iterative decoding threshold has to take into account the respective cases of the interferer being} at the beginning or end of the codeword. Let us denote by  $\idt(\alphal,\mathrm{b})$ and by $\idt(\alphal,\mathrm{e})$ the iterative decoding threshold for a chosen $\alphal$, when the interferer affects the beginning or end of the codeword, respectively.
 We evaluate the following gain function $g$,
 \begin{equation}\label{eq:sec3:gainfct}
 g = \prod_{\ell}\frac{\idt(\alphal,\mathrm{b})}{\interCap(\alphal)}   \cdot \frac{\idt(\alphal,\mathrm{e})}{\interCap(\alphal)}
, \end{equation}
 where $\interCap$ is obtained from the outage capacity expression in~\eqref{eq:TSC}. Note that due to the symmetry constraint in~\eqref{eq:symmetry},  $\idt(\alphal,\mathrm{b}) = \idt(\alphal,\mathrm{e})$ and \eqref{eq:sec3:gainfct} can be simplified.

 \begin{Example}
 	Fix $n_b=11$, $m_b=6$, $p_b=1$, $R_c=1/2$, where the first column is punctured. For the multi-target optimization we impose the symmetry constraint in~\eqref{eq:symmetry}. For complexity reasons, we optimize the threshold for only two different values of $\alpha$, namely $\alpha^{(0)}=6/10$ and $\alpha^{(1)}=9/10$.
 	The multi-target optimization yields\footnote{{The differential evolution algorithm in \cite{Uch14} was executed with the following parameters: crossover probability of $0.6$, population size of $200$, and number of generations of $4000$.}}
 	\begin{equation} \label{eqn_proto_base}
 	\mathbf B_{\mathsf{A}}=\left[
 	\begin{array}{ccccccccccc}
\textcolor{gray}1 & 0 & 0 & 2 & 2 & 0 & 0 & 0 & 0 & 1 & 1\\
\textcolor{gray}2 & 0 & 0 & 0 & 0 & 0 & 1 & 0 & 1 & 1 & 0\\
\textcolor{gray}2 & 0 & 1 & 0 & 1 & 0 & 0 & 0 & 0 & 0 & 1\\
\textcolor{gray}2 & 1 & 0 & 0 & 0 & 0 & 0 & 1 & 0 & 1 & 0\\
\textcolor{gray}2 & 0 & 1 & 1 & 0 & 1 & 0 & 0 & 0 & 0 & 0\\
\textcolor{gray}1 & 1 & 1 & 0 & 0 & 0 & 0 & 2 & 2 & 0 & 0
 	\end{array}
 	\right] .
 	\end{equation}
 \end{Example}

\subsubsection{Raptor-like \ac{LDPC} Code Design} \label{sec:5G}
Protograph-based Raptor-like \ac{LDPC} codes \cite{CDW+11} and belong to the class of rate-compatible \ac{LDPC} codes. The base matrix has the following structure
\begin{equation} \label{eq:raptor}
\mathbf B=\left[
\begin{array}{cc}
	\mathbf B_{\mathsf{pre}} & \mathbf 0 \\
	\mathbf B_{\mathsf{LT}}  & \mathbf I \\
\end{array}
\right]
\end{equation}
where in analogy to Raptor codes $\mathbf B_{\mathsf{pre}}$ and $\mathbf B_{\mathsf{LT}}$ represent the base matrix of the precode and \ac{LT} code respectively. Furhter, $\mathbf I$ is an identity matrix. Owing to their excellent performance, protograph-based Raptor-like \ac{LDPC} codes are adapted in the context of the \ac{5G} standardization for \ac{eMBB}. During the standardization several proposals for base matrices have been made with very similar performance.
We consider one of these \ac{5G} proposals for short blocks \cite{HH17}

\begin{equation}\label{eq:B5G}
\small
\mathbf B_{\mathsf{5G}}=\left[
\begin{array}{cccccccccccccccc|cccccc}
\textcolor{gray}1 & \hspace{0em}\textcolor{gray}0 & \hspace{0em}0 & \hspace{0em}0 & \hspace{0em}1 & \hspace{0em}1 & \hspace{0em}1 & \hspace{0em}0 & \hspace{0em}1 & \hspace{0em}1 & \hspace{0em}1 & \hspace{0em}1 & \hspace{0em}0 & \hspace{0em}0 & \hspace{0em}0 & \hspace{0em}0 & \hspace{0em}0 & \hspace{0em}0 & \hspace{0em}0 & \hspace{0em}0 & \hspace{0em}0 & \hspace{0em}0\\
\textcolor{gray}0 & \hspace{0em}\textcolor{gray}1 & \hspace{0em}0 & \hspace{0em}1 & \hspace{0em}1 & \hspace{0em}1 & \hspace{0em}0 & \hspace{0em}1 & \hspace{0em}0 & \hspace{0em}0 & \hspace{0em}0 & \hspace{0em}1 & \hspace{0em}1 & \hspace{0em}0 & \hspace{0em}0 & \hspace{0em}0 & \hspace{0em}0 & \hspace{0em}0 & \hspace{0em}0 & \hspace{0em}0 & \hspace{0em}0 & \hspace{0em}0\\
\textcolor{gray}1 & \hspace{0em}\textcolor{gray}0 & \hspace{0em}1 & \hspace{0em}0 & \hspace{0em}0 & \hspace{0em}0 & \hspace{0em}1 & \hspace{0em}0 & \hspace{0em}1 & \hspace{0em}1 & \hspace{0em}0 & \hspace{0em}0 & \hspace{0em}1 & \hspace{0em}1 & \hspace{0em}0 & \hspace{0em}0 & \hspace{0em}0 & \hspace{0em}0 & \hspace{0em}0 & \hspace{0em}0 & \hspace{0em}0 & \hspace{0em}0\\
\textcolor{gray}0 & \hspace{0em}\textcolor{gray}1 & \hspace{0em}0 & \hspace{0em}1 & \hspace{0em}0 & \hspace{0em}1 & \hspace{0em}0 & \hspace{0em}1 & \hspace{0em}0 & \hspace{0em}0 & \hspace{0em}1 & \hspace{0em}0 & \hspace{0em}0 & \hspace{0em}1 & \hspace{0em}1 & \hspace{0em}0 & \hspace{0em}0 & \hspace{0em}0 & \hspace{0em}0 & \hspace{0em}0 & \hspace{0em}0 & \hspace{0em}0\\
\textcolor{gray}1 & \hspace{0em}\textcolor{gray}1 & \hspace{0em}1 & \hspace{0em}0 & \hspace{0em}1 & \hspace{0em}0 & \hspace{0em}0 & \hspace{0em}0 & \hspace{0em}1 & \hspace{0em}1 & \hspace{0em}0 & \hspace{0em}0 & \hspace{0em}0 & \hspace{0em}0 & \hspace{0em}1 & \hspace{0em}1 & \hspace{0em}0 & \hspace{0em}0 & \hspace{0em}0 & \hspace{0em}0 & \hspace{0em}0 & \hspace{0em}0\\
\textcolor{gray}1 & \hspace{0em}\textcolor{gray}1 & \hspace{0em}1 & \hspace{0em}1 & \hspace{0em}0 & \hspace{0em}0 & \hspace{0em}1 & \hspace{0em}1 & \hspace{0em}0 & \hspace{0em}0 & \hspace{0em}1 & \hspace{0em}0 & \hspace{0em}0 & \hspace{0em}0 & \hspace{0em}0 & \hspace{0em}1 & \hspace{0em}0 & \hspace{0em}0 & \hspace{0em}0 & \hspace{0em}0 & \hspace{0em}0 & \hspace{0em}0\\\hline
\textcolor{gray}1 & \hspace{0em}\textcolor{gray}1 & \hspace{0em}0 & \hspace{0em}0 & \hspace{0em}0 & \hspace{0em}0 & \hspace{0em}0 & \hspace{0em}0 & \hspace{0em}0 & \hspace{0em}0 & \hspace{0em}0 & \hspace{0em}0 & \hspace{0em}0 & \hspace{0em}0 & \hspace{0em}0 & \hspace{0em}0 & \hspace{0em}1 & \hspace{0em}0 & \hspace{0em}0 & \hspace{0em}0 & \hspace{0em}0 & \hspace{0em}0\\
\textcolor{gray}1 & \hspace{0em}\textcolor{gray}1 & \hspace{0em}0 & \hspace{0em}0 & \hspace{0em}0 & \hspace{0em}0 & \hspace{0em}1 & \hspace{0em}1 & \hspace{0em}0 & \hspace{0em}0 & \hspace{0em}1 & \hspace{0em}0 & \hspace{0em}0 & \hspace{0em}0 & \hspace{0em}0 & \hspace{0em}0 & \hspace{0em}0 & \hspace{0em}1 & \hspace{0em}0 & \hspace{0em}0 & \hspace{0em}0 & \hspace{0em}0\\
\textcolor{gray}0 & \hspace{0em}\textcolor{gray}1 & \hspace{0em}0 & \hspace{0em}0 & \hspace{0em}0 & \hspace{0em}1 & \hspace{0em}1 & \hspace{0em}1 & \hspace{0em}0 & \hspace{0em}0 & \hspace{0em}0 & \hspace{0em}0 & \hspace{0em}0 & \hspace{0em}0 & \hspace{0em}0 & \hspace{0em}0 & \hspace{0em}0 & \hspace{0em}0 & \hspace{0em}1 & \hspace{0em}0 & \hspace{0em}0 & \hspace{0em}0\\
\textcolor{gray}1 & \hspace{0em}\textcolor{gray}0 & \hspace{0em}1 & \hspace{0em}0 & \hspace{0em}0 & \hspace{0em}0 & \hspace{0em}1 & \hspace{0em}1 & \hspace{0em}0 & \hspace{0em}0 & \hspace{0em}0 & \hspace{0em}0 & \hspace{0em}0 & \hspace{0em}0 & \hspace{0em}0 & \hspace{0em}0 & \hspace{0em}0 & \hspace{0em}0 & \hspace{0em}0 & \hspace{0em}1 & \hspace{0em}0 & \hspace{0em}0\\
\textcolor{gray}1 & \hspace{0em}\textcolor{gray}0 & \hspace{0em}0 & \hspace{0em}0 & \hspace{0em}0 & \hspace{0em}0 & \hspace{0em}1 & \hspace{0em}0 & \hspace{0em}1 & \hspace{0em}0 & \hspace{0em}0 & \hspace{0em}0 & \hspace{0em}0 & \hspace{0em}1 & \hspace{0em}0 & \hspace{0em}0 & \hspace{0em}0 & \hspace{0em}0 & \hspace{0em}0 & \hspace{0em}0 & \hspace{0em}1 & \hspace{0em}0\\
\textcolor{gray}1 & \hspace{0em}\textcolor{gray}1 & \hspace{0em}0 & \hspace{0em}0 & \hspace{0em}0 & \hspace{0em}0 & \hspace{0em}0 & \hspace{0em}1 & \hspace{0em}0 & \hspace{0em}0 & \hspace{0em}0 & \hspace{0em}0 & \hspace{0em}0 & \hspace{0em}1 & \hspace{0em}0 & \hspace{0em}0 & \hspace{0em}0 & \hspace{0em}0 & \hspace{0em}0 & \hspace{0em}0 & \hspace{0em}0 & \hspace{0em}1\\
\end{array}
\right]
\end{equation}
where the first two \acp{VN} are punctured. The motivation for this choice is to illustrate that \ac{5G} \ac{eMBB} codes might be  suitable candidates for \ac{RA} applications.

The base matrix in~\eqref{eq:B5G} is not tailored to the specific interference patterns that occur for our asynchronous \ac{RA} protocol. Recall that for our code design we consider interference which is confined to the beginning or end of a codeword. To improve iterative decoding threshold for the base matrix in~\eqref{eq:B5G}, we may permute its columns such that the gain function in~\eqref{eq:sec3:gainfct} is maximized. We evaluate the gain function for two different $\alpha$, i.e., $\alpha^{(0)}=12/20$ and $\alpha^{(1)}=16/20$.

Note that there exist $20!$ permutations for the base matrix $\mathbf B_{\mathsf{5G}}$ (ignoring the punctured columns). To reduce the number of permutations to test, we group \acp{VN} together based on their degrees with the {intuition} that placing a \ac{VN} from a certain group at a position in the protograph will yield a comparable value of the gain function. Therefore, the total number of different {permutations} to be tested reduces to $4200$. The permutation vector yielding the highest value of the gain function is
\begin{align} \label{eq:sec3:perm}
\bm \pi=&\left[0~1~16~4~6~8~10~12~20~2~14~18~19~15~3~21~13~11~9~7~5~17\right] .
\end{align}
We denote the resulting permuted base matrix by $\mathbf B_{\mathsf{5G}}^{\pi}$.

\subsubsection{Asymptotic Results for Gaussian Interference}
\label{sec:ADT}
\begin{figure}
	\centering
	\includegraphics[width=0.9\columnwidth]{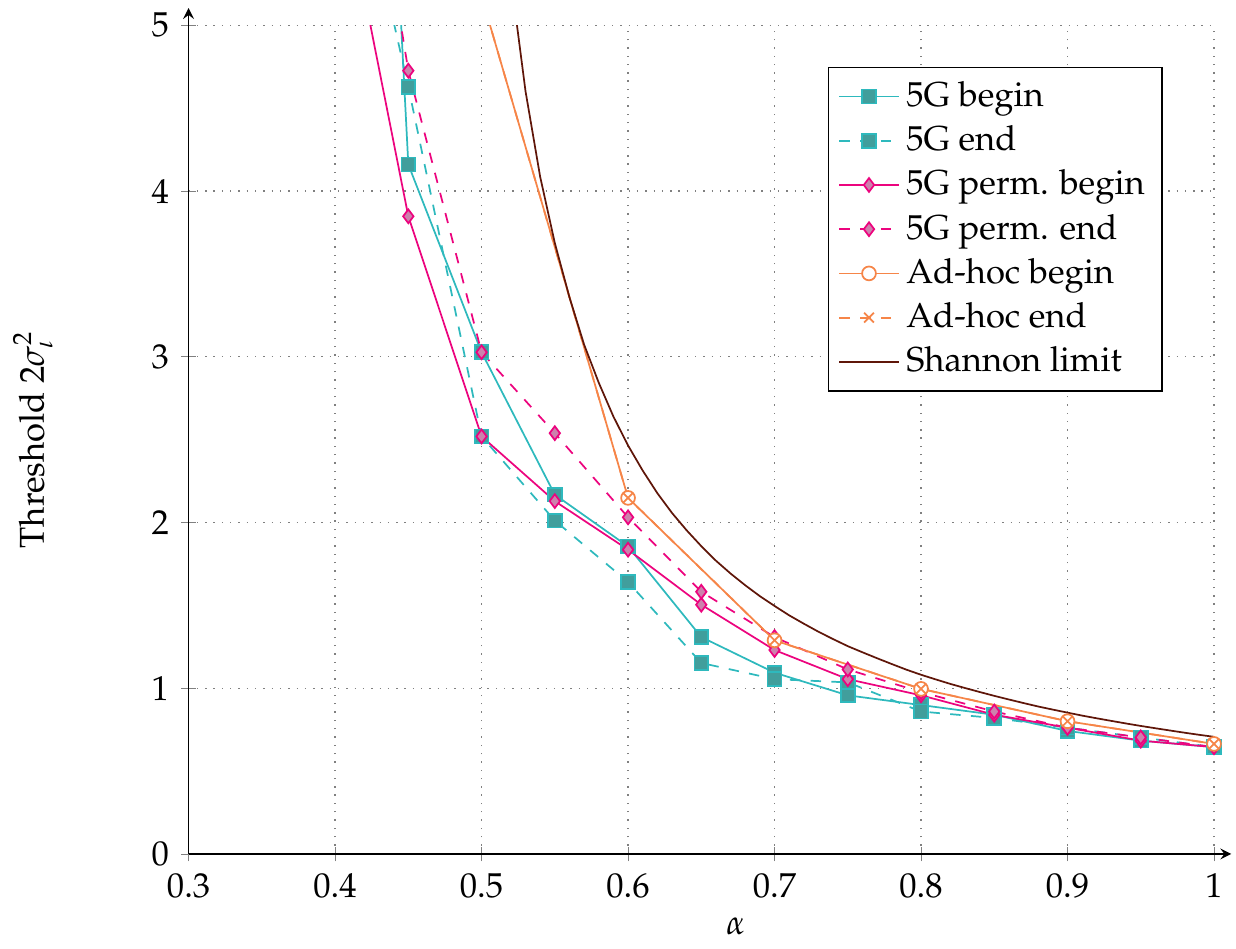}
	\caption{Maximum interference power versus $\alpha$ for different protographs and outage capacity on a QPSK channel with fixed AWGN at $E_s/N_0=6$~dB. The \ac{5G} correseponds to the ensemble with base matrix $\mathbf B_{\mathsf{5G}}$. The \ac{5G} perm. corresponds to the ensemble with base matrix $\mathbf B_{\mathsf{5G}}^{\pi}$. Ad-hoc corresponds to the ensemble with base matrix $\mathbf B_{\mathsf{A}}$. Begin and end represent the iterative decoding threshold with the interference hitting the codeword from the left or the right respectively.}
	\label{fig:asymptotics}
\end{figure}

In Figure~\ref{fig:asymptotics}  we depict iterative decoding thresholds {$\idt$} versus $\alpha$ on the surrogate channel with Gaussian interference for the code ensembles with base matrices $\mathbf B_{\mathsf{A}}$,  $\mathbf B_{\mathsf{5G}}$, and $\mathbf B_{\mathsf{5G}}^{\pi}$. In all three cases the design rate $R_c=1/2$ and $E_s/N_0=6$~dB. We provide thresholds for the cases when the interferer is at the beginning or at the end of a packet. As a reference, we also show $\interCap$ versus $\alpha$ from the outage capacity expression in \eqref{eq:TSC}.

Observe from the figure that the {ad-hoc} code design (base matrix $\mathbf B_{\mathsf{A}}$) performs close to the capacity curve.
Also, owing to the symmetry constraint in~\eqref{eq:symmetry}, the code ensemble shows a the same behavior for the interference being at the beginning or at the end of a codeword. The \ac{5G}-like protograph code ensemble (base matrix $\mathbf B_{\mathsf{5G}}$) performs differently if the interferer is from the left or right. By {permuting} the code (base matrix $\mathbf B_{\mathsf{5G}}^{\pi}$) as described in Section~\ref{sec:5G} the code ensemble performance gets closer to capacity. Further, for $\alpha$ close to one, both \ac{5G}-like and ad-hoc code design perform similarly. This is because here the channel is like a conventional \ac{AWGN} channel for which \ac{5G} codes are known to be among the best \ac{LDPC} codes.

\subsection{Asymptotic Results for a Single Non-Gaussian Interferer}
\label{sec:IM_code_design}

\begin{figure}[!t]
	\centering
	\includegraphics[width=0.9\columnwidth]{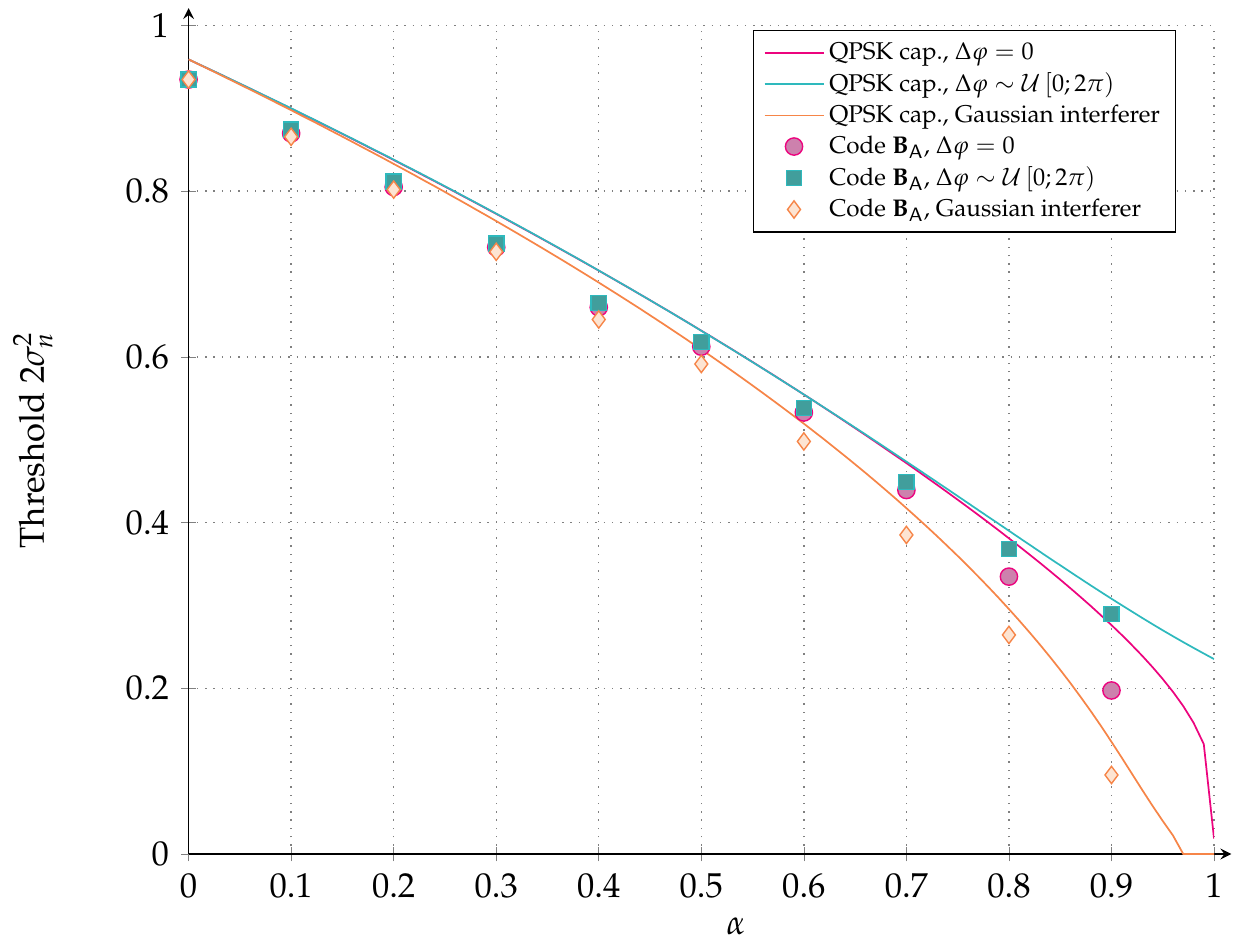}
	\caption{Noise threshold for the code ensamble described by the base matrix $\mathbf B_{\mathsf{A}}$, calculated using quantized density evolution, compared to capacity,  on a QPSK channel with a single equal power interferer as the interferer overlap, $\alpha$, is varied. Three different models for the interferer are considered, a time and phase aligned QPSK signal, a time aligned and random phase QPSK signal and a Gaussian interferer.}
	\label{fig:Threshold_QPSK_CodeBA}
\end{figure}

Although our ad-hoc protograph ensemble was designed for the Gaussian interference model, we show here that its asymptotic performance is also good in this case of non-Gaussian interferers. We consider iterative decoding threshold of our ad-hoc protograph ensemble for two cases:
\begin{itemize}
     \item Symbol-synchronous, phase-aligned, equal-power QPSK interferer, i.e. $\Depoch=0$, $\Dphase=0$, ${\sPw=\iPw=1}$.
     \item Symbol-synchronous, equal-power QPSK interferer with uniform at random phase, i.e. $\Depoch =0$, $\Dphase \sim \mathcal{U}\left[0;2\pi\right)$, ${\sPw=\iPw=1}$.
\end{itemize}
For reference, we also show the results for a single equal-power interferer if that interferer is assumed to be Gaussian.
We determine the iterative decoding threshold of our ad-hoc protograph ensemble for these interference models by applying (quantized) density evolution, making the use of equations~\eqref{eq:llr_qpsk} and \eqref{eq:llr_qpsk_rand}. Since the interferer power is fixed, we define the iterative decoding threshold as the largest noise power per dimension $\noise$ such that error-free transmission is possible. Figure~\ref{fig:Threshold_QPSK_CodeBA} shows the density evolution threshold of ad-hoc protograph ensemble compared to the channel capacity in each of the three interferer models.

As shown in Figure~\ref{fig:Threshold_QPSK_CodeBA}, the QPSK interferer with random phase gives the best threshold, whereas the Gaussian interferer gives the worst threshold. We can explain this by using Figure~\ref{fig:qpsk}. In the absence of noise, the received signal, $\rx(\tm)$ is on a circle around the transmitted signal, $\tx(\tm)$.  To depict $\rx(\tm)$, we can simply draw four circles around the four QPSK constellation points. It can be observed that these circles intersect. However, the probability to pick exactly the intersection point goes to zero. Thus, we can reach a mutual information of two for the QPSK interferer with random phase.
For the phase aligned QPSK interferer, the mutual information between  $\rx(\tm)$ and $\tx(\tm)$ reaches one in the absence of noise. Thus for the interfered part we can get a rate of at most one. Hence the phase aligned QPSK interferer gives a lower threshold than the QPSK interferer with random phase. For the Gaussian interferer case, the mutual information is limited by the QPSK capacity at $E_s/N_0=0$ dB, where the mutual information at 0 dB is 0.96 for the interfered part. Thus the Gaussain interference model show the worst threshold compared to phase aligned QPSK interferer and QPSK intererfer with random phase. 
\section{Numerical Results} \label{sec:results}

For the numerical results we drop the simplified surrogate channel model and focus on two more realistic \ac{RA} channels. The first set of numerical results makes use of a physical layer abstraction where the interference is Gaussian, but not necessarily constant over the $\alpha \ns$ symbols of a modulated codeword. In the second set of numerical results the complete physical layer is implemented and hence the Gaussian assumption on the interference is dropped. While the abstracted physical layer model is a common approach to get first-level insights in the system performance \cite{Clazzer18:ECRA}, (long) physical layer simulations serve as a confirmation of the results. If the identified trends are still present, physical layer results may give evidence on the validity of the surrogate channel model adopted for the code design.

The assumptions common to both set of numerical results are summarized in the following. Users transmit $d=2$ replicas according to the asynchronous \ac{RA} protocol described in Section~\ref{sec:CRA}. Replicas are affected by multiple access interference and by \ac{AWGN} noise, with $E_s/N_0=6$~dB. All replicas are received with the same normalized power $\sPw=1$, thanks to perfect power control. For simplicity, we normalize all duration to the packet length $\pkLen$. The \ac{VF} duration is set to $200 \,\pkLen$, while the receiver decoding window is $\wind= 600 \,\pkLen$. The receiver decoding window shift is set to $\Dwind= 20 \,\pkLen$. 

The performance metric we use to compare the performance of different error correcting codes in the asychronous random access setting is the \ac{PLR} $\plr$, as a function of the channel load $\load$. The \ac{PLR} is the average probability that a user cannot be correctly recovered at the receiver, at the end of the \ac{SIC} process.

\subsection{Abstracted Physical Layer}
\label{sec:apl}

To abstract the physical layer, we make use of  so-called decoding regions \cite{Pulini2013}. Based on a certain interference pattern, we decide whether the decoding of a replica is successful or not, simply by checking whether the corresponding noise plus interference vector falls within the decoding region. An $m$-dimensional decoding region can be seen as an extension of a threshold model to channels with $m$-dimensional channel parameters, such as block fading channels.

\subsubsection{Decoding Region for Random Code Ensembles}
\label{sec:decRegCA}
We  assume that replicas are received with interference whose power may vary over the replica symbols. We resort to a \emph{block interference channel} \cite{McEliece1984}, where a replica experiences various blocks of constant interference. Recall that the interference is assumed to be drawn from a complex white Gaussian process $\mathcal{CN}(0,2\inter)$, with $2\inter$ being the interference power (see also Section \ref{sec:gim}). 

Let us call $m'$ the maximum number of interferers affecting a replica. Let us denote by
$m$, $1 \leq m\leq m'+1$, the number of \emph{different} interference plus noise levels that are present over the considered replica. Define 
\begin{equation}\label{eq:noiseplusinter}
\BnPI=\Big[\underbrace{\noise+\frac{1}{2}}_{\nPI_1},\ldots,\underbrace{\noise+\frac{m}{2}}_{\nPI_m}\Big]
\end{equation}
as the ordered interference plus noise vector which contains the  $m$ different noise plus interference levels in the replica. Let $\bm{\alpha}=\left[\alpha_1,\ldots,\alpha_m\right]$, where  $\alpha_j$, $j\in\{ 1,\ldots, m \}$, is the fraction of a replica subject to the interference plus noise of power $\left(\noise+\frac{j}{2}\right)$ per dimension. Clearly, $\sum_{j=1}^m \alpha_j \leq 1$ and $\alpha_j \in(0,1]$. Recalling that $\Ca\left(\nPI\right)$ is the \ac{QPSK} constraint \ac{AWGN} channel capacity for a given $\nPI$, the outage capacity $\Capa$ under the Gaussian interference assumption is
\begin{equation}
\label{eq:TSC_full}
\Capa(\bm{\alpha}, \noise, \BnPI) = \left(1-\sum_{j=1}^{m}\alpha_j\right) \,\Ca\left(\frac{1}{2\noise}\right) + \alpha_1 \, \Ca\left(\frac{1}{2\nPI_1}\right)+\ldots+ \alpha_m \, \Ca\left(\frac{1}{2\nPI_m}\right).
\end{equation}
For a fixed transmission rate $\rate$, we define the decoding region $\rvDec$ \cite{Boutros2005,Pulini2013} as
\begin{align}\label{eq:dec_reg_full}
\rvDec &= \Bigg\{ \BnPI \in \mathbb{R}_{+}^{m}, \bm{\alpha} \in \left(0,1\right]^{m} \mid \sum_{j=1}^{m} \alpha_j \leq 1 \wedge \rate < \Capa(\BnPI, \noise, \bm{\alpha}) \Bigg\} .
\end{align}
Every $m'$-user collision resulting in $(\bm{\alpha}, \noise, \BnPI) \in \rvDec$ can be resolved. 

A possibility  to numerically evaluate the performance of a random code ensemble which achieves a fraction $\beta$ of the outage capacity with $\capBO<1$, is to compute the decoding region $\rvDec'\subset \rvDec$ as 
\begin{align}\label{eq:dec_reg_full_2}
\rvDec' &= \Bigg\{ \BnPI \in \mathbb{R}_{+}^{m}, \bm{\alpha} \in \left(0,1\right]^{m} \mid \sum_{j=1}^{m} \alpha_j \leq 1 \wedge \rate < \capBO\,\Capa(\BnPI, \noise, \bm{\alpha}) \Bigg\} .
\end{align}

For the abstracted physical layer simulations we follow the steps from the literature \cite{Pulini2013}. We generate realizations of packet collisions for a certain channel load $\load$. For each of the $\ns$ codeword symbols of a replica, we compute an instantaneous noise plus interference power. We assume that for each codeword symbol we have knowledge of the number of interferers due to ideal detection. Thus, we  group the $\ns$ instantaneous noise plus interference power values into $m$ blocks, sort them according to the number of interferers, average over the noise plus interference power in every block to finally obtain the ordered interference plus noise vector  $\BnPI$ in~\eqref{eq:noiseplusinter}. Successful decoding is declared if  $(\bm{\alpha}, \noise, \BnPI) \in \rvDec$ (or $\in \rvDec'$). 

\subsubsection{Decoding Region for LDPC Code Ensembles}
\label{sec:decReg}

For \ac{LDPC} protograph ensembles the decoding region is computed slightly differently from~\eqref{eq:dec_reg_full}. Since for structured code ensembles not only the power and fraction, but also the position of the interferer is important, we introduce a length-$n_b$ protograph noise plus interference vector $\BnPIo=\left[ \left(\noise+\inter\right)_{0},\ldots, \left(\noise+\inter\right)_{n_b-1}    \right]$. An element $\left(\noise+\inter\right)_{j}$ corresponds to the specific signal plus interference level for a \ac{VN} of type $j$. Then, the decoding region in~\eqref{eq:dec_reg_full} can be restated as
\begin{align}
\label{eq:dec_cond_full_code}
\rvDec'' &= \Bigg\{\BnPIo \in \mathbb{R}^{n_b}_{+} \mid \IAPP(j) \rightarrow 1\,\forall j\in\{0, \ldots, n_b-1\}\Bigg\}.
\end{align}

The  signal plus interference level $\left(\noise+\inter\right)_{j}$ for a \ac{VN} of type $j$ is determined as follows. For a certain channel load $\load$, realizations of packet collisions are generated. For each of them one may determine an instantaneous bit-wise noise plus interference power for each of the $n$ codeword bits. Then, for each of the $n_b-p_b$ unpunctured \ac{VN} types in the protograph we determine an average noise plus interference level by averaging over $n/(n_b-p_b)$ subsequent values of the instantaneous bit-wise noise plus interference powers yielding $\BnPIo$. Decoding is  successful only if $\BnPIo \in \rvDec''$.

\subsubsection{Simulation Results}
\label{sec:results_mac}
\begin{figure}[!t]
\centering
\includegraphics[width=0.9\columnwidth]{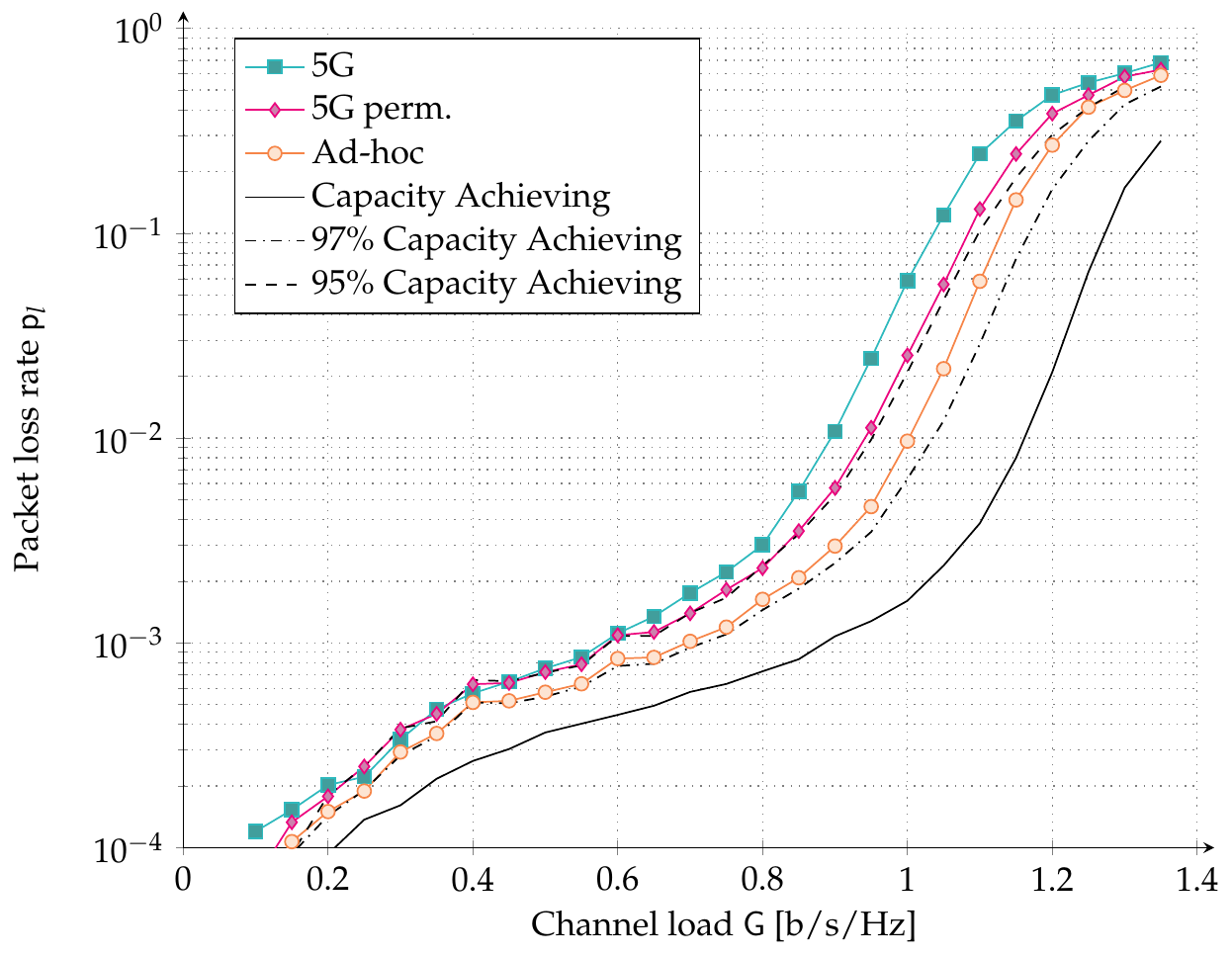}
\caption{{\ac{PLR} vs.\ channel load $\load$ for the \ac{5G}, permuted \ac{5G}, and ad-hoc \ac{LDPC} protograph code ensembles for asynchronous \ac{RA} setting with abstracted physical layer. The \ac{5G} correseponds to the ensemble with base matrix $\mathbf B_{\mathsf{5G}}$. The \ac{5G} perm. corresponds to the ensemble with base matrix $\mathbf B_{\mathsf{5G}}^{\pi}$. Ad-hoc corresponds to the ensemble with base matrix $\mathbf B_{\mathsf{A}}$.}}
\label{fig:PER_MAC}
\end{figure}
The \ac{PLR} performance assuming the three different protograph code ensembles from Section~\ref{sec:code_design} is shown in Figure~\ref{fig:PER_MAC}. For reference, the performance for a capacity achieving code ensemble, one which achieves $97\%$ of capacity and one which achieves $95\%$ of capacity are also shown (solid black, dot-dashed black and dashed black curves, respectively). In the entire channel load range, the ad-hoc design (base matrix $\mathbf B_{\mathsf{A}}$) visibly outperforms both \ac{5G}-based solutions (base matrices $\mathbf B_{\mathsf{5G}}$ and $\mathbf B_{\mathsf{5G}}^{\pi}$). The simple yet insightful observation that the beginning and the end of a packet shall be strongly protected by the channel code, leads to a beneficial performance gain also when the overall multiple access interference comes into play. In particular, for a target \ac{PLR} of $10^{-3}$, the channel load supported can be extended from $0.6$~[b/s/Hz] of the \ac{5G}-based solutions to $0.7$~[b/s/Hz] of the ad-hoc \ac{LDPC} design, resulting in a $0.1$~[b/s/Hz] or $17\%$ gain. Similarly, at \ac{PLR} of $10^{-2}$, the channel load supported can be extended from $0.9$~[b/s/Hz] of the \ac{5G}-based solutions to $1.0$~[b/s/Hz] of the ad-hoc \ac{LDPC} design, resulting in a $0.1$~[b/s/Hz] or $11\%$ gain. For a fixed channel load operating point, the gain is even more remarkable. 

Finally, observe that there is still a visible gap to  capacity achieving code ensembles. One may erroneously conclude that the constant interfering power assumption of the surrogate channel model in Section~\ref{sec:surrogate} is inaccurate, yielding a design that is penalized on the Gaussian interference channel with  varying interference powers. However, we found that even a small loss in error correction performance may have a big effect on the \ac{PLR}. To underpin this observation, consider a code ensemble that achieves $95\%$ of the outage capacity. Observe that this code ensemble performs similarly to the permuted \ac{5G}-like design and worse than our ad-hoc design for the surrogate channel. A possible reason for the drastic impact of the code performance on the asynchronous random access \ac{PLR} might due to the \ac{SIC}: in case a  replica cannot be correctly decoded due to the sub-optimality of the error correcting code, it may arrest the \ac{SIC} process and cause multiple packet losses. Consider a code ensemble that achieves $97\%$ of the outage capacity instead of $95\%$. Observe from the figure that such a minor improvement on the code design can result in gains of up to $0.1$~[b/s/Hz].

In order to confirm the trends identified with the abstracted physical layer and to overcome its inherent limitations, in the following Section we will resort to simulate the physical layer performance of the proposed \ac{LDPC} codes. 

\subsection{Finite-Length Physical Layer Simulations with the Designed \ac{LDPC} Codes}
\label{sec:results_phy}
\begin{figure}[!t]
\centering
\includegraphics[width=0.9\columnwidth]{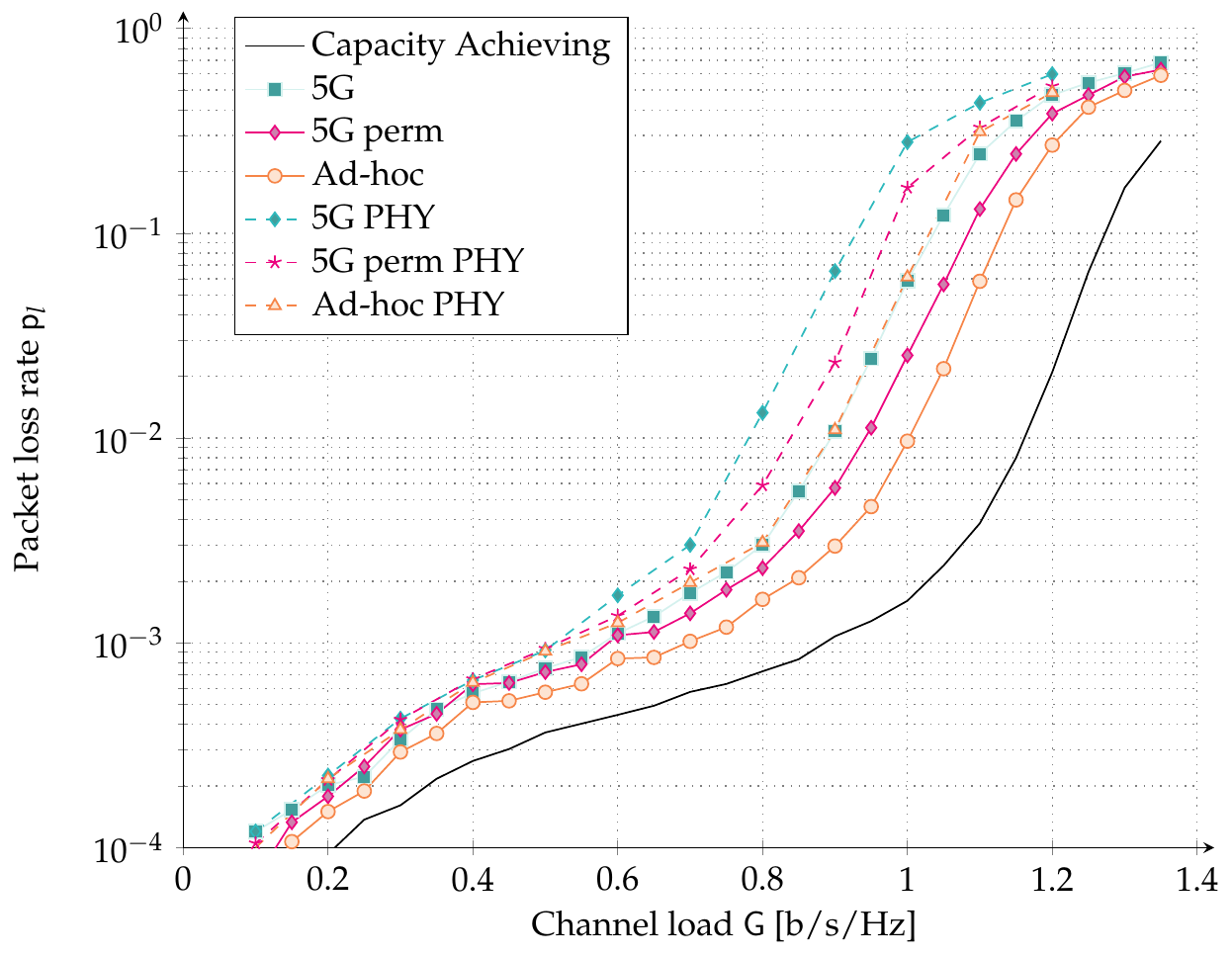}
\caption{Dashed: \ac{PLR} vs.\ channel load $\load$ for the asynchronous \ac{RA} protocol with replicas protected by different $(960,480)$ \ac{LDPC} codes, as a result of physical layer simulations. Solid: curves from Figure~\ref{fig:PER_MAC} assuming abstracted physical layer. The \ac{5G} correseponds to the ensemble with base matrix $\mathbf B_{\mathsf{5G}}$. The \ac{5G} perm. corresponds to the ensemble with base matrix $\mathbf B_{\mathsf{5G}}^{\pi}$. Ad-hoc corresponds to the ensemble with base matrix $\mathbf B_{\mathsf{A}}$.}
\label{fig:PER_PHY}
\end{figure}
We consider here a $(960,480)$ \ac{LDPC} codes obtained from the base matrices in  Section~\ref{sec:code_design}. The code parameters are chosen to fit in the context of short packet communications.\footnote{Packets in the order of hundreds of information bits are typical of \ac{M2M} applications. An example is the early data transmission procedure available in release 15 of 3GPP, where terminals are allowed to piggyback data on message $3$ of the \ac{PRACH} with sizes between $328$ and $1000$ information bits \cite{5g_release15}.}  Each users selects a codeword uniformly at random from the codebook. The codeword is \ac{QPSK} modulated, and two instances (replicas) are transmitted. Each of the  modulated replicas are sent over the asynchronous \ac{RA} channel undergoing the protocol rules as described in Section~\ref{sec:system_model}. Ideal channel estimation is assumed, so that the receiver can perfectly compensate for the phase offset. After channel estimation, soft-\ac{LLR} values are computed based on the perfect interference power knowledge. Note that the interference contribution here is  the superposition of possibly multiple \ac{QPSK} modulated, equal-power signals, with random phase offsets, corresponding to all replicas that are concurrently received. It is important to stress that this assumption departs from the Gaussian interference assumption, especially when the number of interfering replicas is small. A standard belief propagation \ac{LDPC} decoder with a maximum of $50$ decoding iterations is used to counteract the effect of noise and multiple access interference. If the decoder is successful, the replica is removed from the received signal together with its twin, so that no residual interference power is left after cancellation.

The dashed curves in Figure~\ref{fig:PER_PHY} show the \ac{PLR} performance versus channel load of the asynchronous \ac{RA} scheme with replicas protected with a code described by the base matrix $\mathbf B_{\mathsf{A}}$ (named Ad-hoc PHY), $\mathbf B_{\mathsf{5G}}^{\pi}$ (named 5G perm PHY), or $\mathbf B_{\mathsf{5G}}$ (named 5G PHY) respectively. The solid curves are resumed from Figure~\ref{fig:PER_MAC} and depict results for the abstracted physical layer. Observe that the relative gap between the ad-hoc \ac{LDPC} design, permuted \ac{5G} design, as well as the \ac{5G} design is preserved, since all three codes suffer from a similar degradation due to finite length effects. For the ad-hoc design, at a target \ac{PLR} of $10^{-3}$, the supported channel load is reduced from $0.7$~[b/s/Hz] to $\sim0.6$~[b/s/Hz]. Similarly at $10^{-2}$ the achieved channel load becomes $0.9$~[b/s/Hz] compared to $1$~[b/s/Hz] when the abstract physical layer is considered. In line with the ad-hoc design, also the \ac{5G} code suffers from a degradation of $\sim0.1$~[b/s/Hz] in the channel load region of interest. 

Interestingly, although the ad-hoc \ac{LDPC} code is designed for a surrogate channel, both the abstract physical layer and the full physical layer simulations confirm the beneficial impact on the asynchronous random access protocol. The abstract physical layer model is able to hinge the performance tendencies that are subject to a penalty when one considers the precise physical layer. Such penalty mostly originates from finite-length effects of the channel code. The assumption of Gaussianity for the multiple access interference, as well as constant interfering power appears to be reasonable and thus can be exploited to simplify the channel code design.			
\section{Conclusions}
This work presents  protograph \ac{LDPC} code designs for asynchronous \ac{RA} with \ac{SIC}. For the code design we make use of a simplified surrogate channel model which assumes Gaussian interference and constant interfering power over the interfered part of a packet. Considering more realistic \ac{RA} channels, we show that the  proposed designs for the surrogate channel perform close to theoretical limits. Gains of  $17$\%, in supported channel load at a packet loss rate of $10^{-2}$ are observed w.r.t.\ an off-the-shelf code. In contrast to literature where the physical layer code performance is often abstracted by means of decoding thresholds or decoding regions, physical layer LDPC code simulations for short blocks are performed, yielding a more realistic estimate of the achievable supported channel load (for a fixed target packet error rate). We find that abstracting the physical layer overestimates the performance for our setup by approximately $10$\% in supported channel load, at a packet loss rate of $10^{-2}$.

%%%%%%%%%%%%%%%%%%%%%%%%%%%%%%%%%%%%%%%%%%
\section*{Acknowledgments}
The authors would like to thank Nina Grosheva for her support in the design of the asnychronous medium access simulation environment during its early stages.

\ifCLASSOPTIONcaptionsoff
  \newpage
\fi

\bibliographystyle{IEEEtran}
\bibliography{IEEEabrv,References}
%%%%%%%%%%%%%%%%%%%%%%%%%%%%%%%%%%%%%%%%%%
\end{document}